\documentclass[journal]{IEEEtran}

\makeatletter
\def\ps@headings{
\def\@oddhead{\mbox{}\scriptsize\rightmark \hfil \thepage}
\def\@evenhead{\scriptsize\thepage \hfil \leftmark\mbox{}}
\def\@oddfoot{}
\def\@evenfoot{}}
\makeatother 

\IEEEoverridecommandlockouts

\usepackage{amsfonts}
\usepackage[dvips]{graphicx}
\usepackage[caption=false,font=normalsize,labelfont=sf,textfont=sf]{subfig}
\usepackage{times}
\usepackage{cite}
\usepackage{lettrine}
\usepackage{amsmath}
\usepackage{amsmath}
\allowdisplaybreaks[4]
\usepackage{array}
\usepackage{amssymb}

\usepackage{stfloats}
\usepackage{slashbox}
\usepackage{graphicx}
\usepackage{footnote}
\usepackage{booktabs}
\usepackage{array}
\usepackage{algorithmic}
\usepackage{algorithm}
\usepackage{subeqnarray}
\usepackage{cases}
\usepackage{threeparttable}
\usepackage{color}
\usepackage{bm}
\usepackage{bbm}

\usepackage{capt-of} 
\usepackage{tabularx}
\usepackage{multirow}
\DeclareMathOperator*{\argmax}{argmax}
\DeclareMathOperator*{\argmin}{argmin}

\newtheorem{theorem}{\underline{Theorem}}[section]
\newtheorem{lemma}{\underline{Lemma}}[section]
\newtheorem{corollary}{\underline{Corollary}}[section]
\newtheorem{proposition}{Proposition}[section]

\newtheorem{remark}{\underline{Remark}}[section]

\begin{document}
	\bibliographystyle{IEEEtran}

	\title{
		CFLIT: Coexisting Federated Learning and Information Transfer
	}

	\IEEEoverridecommandlockouts
	\author{
		Zehong Lin,~\IEEEmembership{Member,~IEEE}, Hang Liu,~\IEEEmembership{Member,~IEEE}, and Ying-Jun Angela Zhang,~\IEEEmembership{Fellow,~IEEE}
		
		\thanks{Zehong Lin was with the Department of Information Engineering, The Chinese University of Hong Kong, Hong Kong. He is now with the Department of Electronic and Computer Engineering, The Hong Kong University of Science and Technology, Hong Kong (e-mail: eezhlin@ust.hk).
			
		Hang Liu was with the Department of Information Engineering, The Chinese University of Hong Kong, Hong Kong. He is now with the Department of Electrical and Computer Engineering, Cornell Tech, Cornell University, New York, NY 10044 USA (e-mail: hl2382@cornell.edu).
			
		Ying-Jun Angela Zhang is with the Department of Information Engineering, The Chinese University of Hong Kong, Hong Kong (e-mail: yjzhang@ie.cuhk.edu.hk).
		}
	}
	\maketitle

	\begin{abstract}
		Future wireless networks are expected to support diverse mobile services, including artificial intelligence (AI) services and ubiquitous data transmissions. Federated learning (FL), as a revolutionary learning approach, enables collaborative AI model training across distributed mobile edge devices. By exploiting the superposition property of multiple-access channels, over-the-air computation allows concurrent model uploading from massive devices over the same radio resources, and thus significantly reduces the communication cost of FL. In this paper, we study the \emph{coexistence} of over-the-air FL and traditional information transfer (IT) in a mobile edge network, where an access point (AP) coordinates a set of devices for over-the-air FL and serves multiple devices for information transfer in the meantime. We propose a \emph{coexisting federated learning and information transfer} (\emph{CFLIT}) communication framework, where the FL and IT devices share the wireless spectrum in an orthogonal frequency division multiplexing (OFDM) system. Under this framework, we aim to maximize the IT data rate and guarantee a given FL convergence performance by optimizing the long-term radio resource allocation. A key challenge that limits the spectrum efficiency of the coexisting system lies in the large overhead incurred by frequent communication between the server and edge devices for FL model aggregation. To address the challenge, we rigorously analyze the impact of the computation-to-communication ratio on the convergence of over-the-air FL in wireless fading channels. The analysis reveals the existence of an optimal computation-to-communication ratio that minimizes the amount of radio resources needed for over-the-air FL to converge to a given error tolerance. Based on the analysis, we propose a low-complexity online algorithm to jointly optimize the radio resource allocation for both the FL devices and IT devices. We further derive an analytical expression of the achievable data rate of IT users. Extensive numerical simulations verify the superior performance of the proposed design for the coexistence of FL and IT devices in wireless cellular systems.	
	\end{abstract}


	\section{Introduction}	
		\IEEEPARstart{T}{he} proliferation of mobile artificial intelligence (AI) services, such as autonomous driving and intelligent personal assistants, triggers the emergence of a new paradigm called \emph{edge intelligence} \cite{zhou2019edge, park2019wireless, letaief2019roadmap}, which migrates the training and inference processes of AI models from central cloud servers to the edge of wireless networks, i.e., edge servers and mobile devices. Federated learning (FL) \cite{mcmahan2017communication} has been widely perceived as a promising edge learning framework that allows mobile devices to collaboratively train a shared AI model without disclosing the raw data. Specifically, mobile devices download the global model from an edge server for local model training and upload the updated local models to the edge server. The server then updates the global model by aggregating a linear combination of local models. To relieve the communication burden of FL model uploading and aggregation, over-the-air computation \cite{nazer2007computation} has been introduced into the FL model aggregation process \cite{yang2020federated}. By exploiting the superposition property of multiple-access channels, over-the-air model aggregation allows massive devices to simultaneously transmit their local models using the same channel and coherently aligns the local models at the server. The required bandwidth and communication latency for over-the-air computation are independent of the number of devices, thus significantly reducing the communication cost for FL model aggregation.
		
		Extensive studies have been conducted to design over-the-air FL systems in wireless networks \cite{yang2021privacy, yang2020federated,  liu2020reconfigurable, lin2021relay, cao2021optimized, zhang2021gradient, xu2021learning, liu2020privacy}. Most prior work, however, considers over-the-air computation over narrow-band systems and is not applicable to modern broadband wireless networks. To facilitate broadband implementation, the recent work in \cite{zhu2019broadband} proposed a broadband over-the-air model aggregation approach in orthogonal frequency division multiplexing (OFDM) systems. Nonetheless, Ref. \cite{zhu2019broadband} ignored the coexistence of over-the-air FL with traditional information transfer (IT) services. Note that future wireless networks are required to support multiple devices with different data communication demands at the same time \cite{saad2019vision, tataria20216g, letaief2021edge}. For instance, FL devices need to frequently communicate model information with an access point (AP), which in the meanwhile serves another set of devices for data streaming. Therefore, it is of great importance to design a coexisting system to support the coexistence of over-the-air FL and IT in wireless networks. This motivates the work in this paper.
		
	    Designing the above coexisting system is a challenging task. Although over-the-air model aggregation greatly reduces the bandwidth cost compared with traditional FL, uploading high-dimensional model parameters to the edge server iteratively until model convergence still leads to an extremely high bandwidth cost. Such a high bandwidth consumption leaves very little spectrum for IT devices to transmit their data traffic. Therefore, minimizing the bandwidth consumption of over-the-air FL is a critical problem in the design of the coexisting system. Note that the total bandwidth consumption is sensitive to the total number of communication rounds, which is critically determined by the design of the over-the-air FL training process. Consequently, we shall carefully study the FL mechanism design to minimize the bandwidth consumption needed for convergence.
	
		There are two main research directions towards minimizing the bandwidth consumption of FL in the literature: one is to reduce the communication bandwidth cost per communication round and the other is to reduce the total number of communication rounds for model update. On one hand, the communication cost in each round can be reduced by model compression \cite{lin2017deep}. Existing approaches for model compression are divided into two categories: quantization \cite{alistarh2018qsgd} and sparsification \cite{stich2018sparsified}. In the former, local models are quantized into finite-bit digits to reduce the amount of data in model uploading \cite{bernstein2018signsgd, zhu2020one}. In the latter, local gradients are transformed to low-dimensional vectors at the transmitters by exploiting the gradient sparsity and recovered at the server by compressed sensing \cite{amiri2020machine, amiri2020federated, fan2021temporal}. On the other hand, the total number of communication rounds can be reduced by carefully selecting the computation-to-communication ratio. For example, \cite{mcmahan2017communication, stich2018local, li2020federated} have reported that the number of model updates computed locally before uploading the local model parameters to the server for model aggregation critically determines the total number of communication rounds needed for convergence. Specifically, a smaller number of local updates (i.e., a lower computation-to-communication ratio) means more frequent model aggregation, and thus a higher communication cost. Meanwhile, a larger number of local updates (i.e., a higher computation-to-communication ratio) leads to infrequent model aggregation, which inevitably slows down the learning convergence especially when the data distribution across devices is heterogeneous. Inspired by this observation, the authors in \cite{li2019convergence} optimized the number of local updates to minimize the required number of FL communication rounds. Moreover, Refs. \cite{wang2019adaptive, amiri2021convergence} extended the result in \cite{li2019convergence} by taking the wireless network characteristics into account.
		
		Although the current work in \cite{amiri2020machine, amiri2020federated, fan2021temporal} has demonstrated the effectiveness of model compression on reducing the communication cost for over-the-air model aggregation, the number of model parameters after compression is still very large. For instance, the famous deep learning model AlexNet in  \cite{krizhevsky2012imagenet} consists of more than 60 million parameters. Even the state-of-the-art gradient sparsification approach has compressed the transmitted local gradients with compression ratio of  $10\%$ \cite{fan2021temporal}, transmitting the sparsified model to the server still consumes a large amount of radio resources in each round. In light of this, reducing the required number of model communication rounds is critical to save the bandwidth cost. The prior work on the optimization of computation-to-communication ratio \cite{li2019convergence, wang2019adaptive, amiri2021convergence}, however, considers conventional orthogonal transmission and is not applicable to over-the-air FL. For over-the-air FL, the wireless channel fading inevitably induces signal misalignment in model aggregation, which brings a considerable error in the aggregated global gradient. Such an error is coupled with the effect of local model updating during the training process and slows down the learning convergence rate \cite{liu2020reconfigurable}. Therefore, it is necessary to investigate the coupled impact of the model aggregation error and the number of local updates on the bandwidth consumption needed for the convergence of over-the-air FL. Nonetheless, this important study is largely overlooked in the literature.
	
	In this paper, we study a \emph{coexisting federated learning and information transfer} (\emph{CFLIT}) system in a single-cell mobile edge network consisting of a single AP, a set of FL devices, and a set of IT devices. The AP coordinates the FL devices for over-the-air FL and simultaneously serves the IT devices for uplink IT using the same uplink spectrum. To minimize the bandwidth cost for FL, we first analyze the convergence behavior of over-the-air FL as a function of local model updates and the total number of communication rounds in a wireless fading channel. Based on the analysis, we optimize the key parameters, i.e., the number of local updates and the number of FL communication rounds, to minimize the radio resources needed for over-the-air model aggregation. Then, we formulate a radio resource allocation problem to simultaneously maximize the average IT data rate and guarantee the convergence of over-the-air FL in a given time horizon. The lack of non-causal information of future channel state makes it difficult to jointly optimize the long-term radio resource allocation for both the FL and IT devices with an offline scheme. To overcome this difficulty, we propose a low-complexity online algorithm to make real-time allocation decisions based on the current observation of channel state information (CSI). To the best of our knowledge, this is the first paper that studies the joint communication and learning design in a wireless network to support the coexistence of over-the-air FL and IT. The main contributions of this paper are summarized as follows.
		\begin{itemize}
			\item \emph{Bandwidth cost minimization for over-the-air FL:} We theoretically analyze the achievable training loss of over-the-air FL under mild assumptions to characterize the impact of the computation-to-communication ratio on the over-the-air FL convergence. We then derive a closed-form expression of the optimal computation-to-communication ratio that yields the minimum number of communication rounds. The analysis provides a useful guideline on how to optimize the key parameters of FL for minimizing the bandwidth cost for over-the-air model aggregation.
			
			\item \emph{CFLIT communication framework:} We develop a novel CFLIT communication framework to coordinate both over-the-air model aggregation and uplink data streaming. Specifically, the FL devices and the IT devices simultaneously transmit local model updates and general data streams, respectively, to the AP over time-varying broadband channels. We formulate a CFLIT radio resource allocation problem to maximize the average IT data rate and at the same time ensure FL convergence in a given duration. To minimize the radio resource consumption for FL convergence guarantee, the transceiver design and learning parameters are jointly taken into account.
		
			\item \emph{Low-complexity online algorithm for radio resource allocation:} We propose an efficient threshold-based online subcarrier allocation algorithm for CFLIT by mimicking the offline optimal solution. Unlike the offline scheme, the proposed algorithm does not require the non-causal CSI of the entire time horizon. Instead, it relies on the instantaneous channel gains of IT devices in the current symbol duration only and allocates subcarriers based on an optimized threshold. Moreover, we provide theoretical performance analysis to prove the rate improvement of the proposed online algorithm.
		\end{itemize}
	
	Simulation results verify our theoretical analysis and show that the proposed online algorithm performs very close to the offline optimal solution. Moreover, the proposed design achieves a better FL performance and a higher average IT data rate compared with the baseline approaches.
	
	The rest of this paper is organized as follows. In Section II, we describe the CFLIT system model and the problem formulation. In Section III, we present the convergence analysis for over-the-air FL. Based on this, we derive analytical expressions to optimize the number of local updates and the number of communication rounds. In Section IV, we propose an efficient online algorithm to optimize the subcarrier allocation and provide theoretical performance analysis. In Section V, we evaluate the proposed design via extensive simulations. Finally, we conclude the paper in Section VI.
	
	\emph{Notations}: We use regular letters, boldface lower-case letters, boldface upper-case letters, and calligraphy letters to denote scalars, vectors, matrices, and sets, respectively. The real and complex domains are denoted by $\mathbb{R}$ and $\mathbb{C}$, respectively. We use $\mathcal{N}(\mu, \sigma^2)$ and $\mathcal{CN}(\mu, \sigma^2)$ to denote the real and the circularly-symmetric complex Gaussian distributions with mean $\mu$ and variance $\sigma^2$, respectively. We use $x_i$ or $x[i]$ to denote the $i$-th entry of vector $\mathbf{x}$, $x_{ij}$ to denote the $(i, j)$-th entry of matrix $\mathbf{X}$, $|\mathcal{S}|$ to denote the cardinality of set $\mathcal{S}$, and $\mathbb{E}[\cdot]$ to denote the expectation operator.

	\section{System Model and Problem Formulation}	
	
	\subsection{CFLIT System Model}
	We consider a single-cell CFLIT system, where $K$ FL devices, indexed by $\mathcal{K} = \{1, \cdots, K\}$, collaboratively train an AI model by exchanging model information with an AP. At the same time, $N$ IT devices, indexed by $\mathcal{N} = \{1, \cdots, N\}$, transmit information-bearing signals to the AP through the uplink channels. All the nodes are equipped with single antenna. As shown in \cite{konevcny2016federated}, the uplink model uploading and aggregation incur a high bandwidth cost and act as the communication bottleneck of FL. To significantly reduce the communication cost for FL, we adopt the over-the-air computation technique to accelerate the uplink model aggregation.\footnote{Over-the-air computation requires synchronization of the transmitters. For simplicity, we assume in this paper that the local model information transmitted from the FL devices is perfectly synchronized, which can be achieved by the state-of-the-art synchronization mechanism; see, e.g., \cite{guo2021over}. We refer to \cite{guo2021over} for more technical details on over-the-air signal synchronization.} By employing OFDM modulation, the total uplink bandwidth is divided into $M$ orthogonal OFDM subcarriers, indexed by $\mathcal{M} = \{1, \cdots, M\}$. The AP is in charge of allocating subcarriers to support the over-the-air model aggregation of the FL devices and the data transmissions from the IT devices. We consider a time horizon consisting of $S$ OFDM symbols, where $S$ represents the allowed time for accomplishing the FL task and is determined by the specific application. The total number of available OFDM resource blocks (RBs) for accomplishing over-the-air FL and IT is $MS$. To avoid large inter-user interference, each subcarrier is allocated to either over-the-air model aggregation or at most one IT device.
	
	We assume a block fading channel model, where the channel coefficients between the devices and the AP evolve independently over each coherence block following the independent and identically distributed (i.i.d.) Gaussian distribution of $\mathcal{CN}(0, 1)$. Suppose that the considered time interval of $S$ OFDM symbols is much larger than the coherence block length. We need to consider long-term dynamic RB allocation over multiple blocks with varying channel coefficients. We consider a training-based channel estimation protocol where a number of time slots at the beginning of each coherence block is used to transmit pilots for uplink channel estimation. We ignore the overhead and error in channel estimation\footnote{We note that the channel estimation overhead and error have a negative impact on system performance. The overhead of channel estimation reduces the available number of RBs, while the estimation error potentially leads to inaccurate RB allocation, making the solution sub-optimal. To mitigate these effects, an efficient channel estimation scheme is needed, which, however, is beyond the scope of this paper.} and assume that only the CSI of the current coherence block is available.
	
	\begin{figure*}[t]
		\centering
		\includegraphics[scale=0.42]{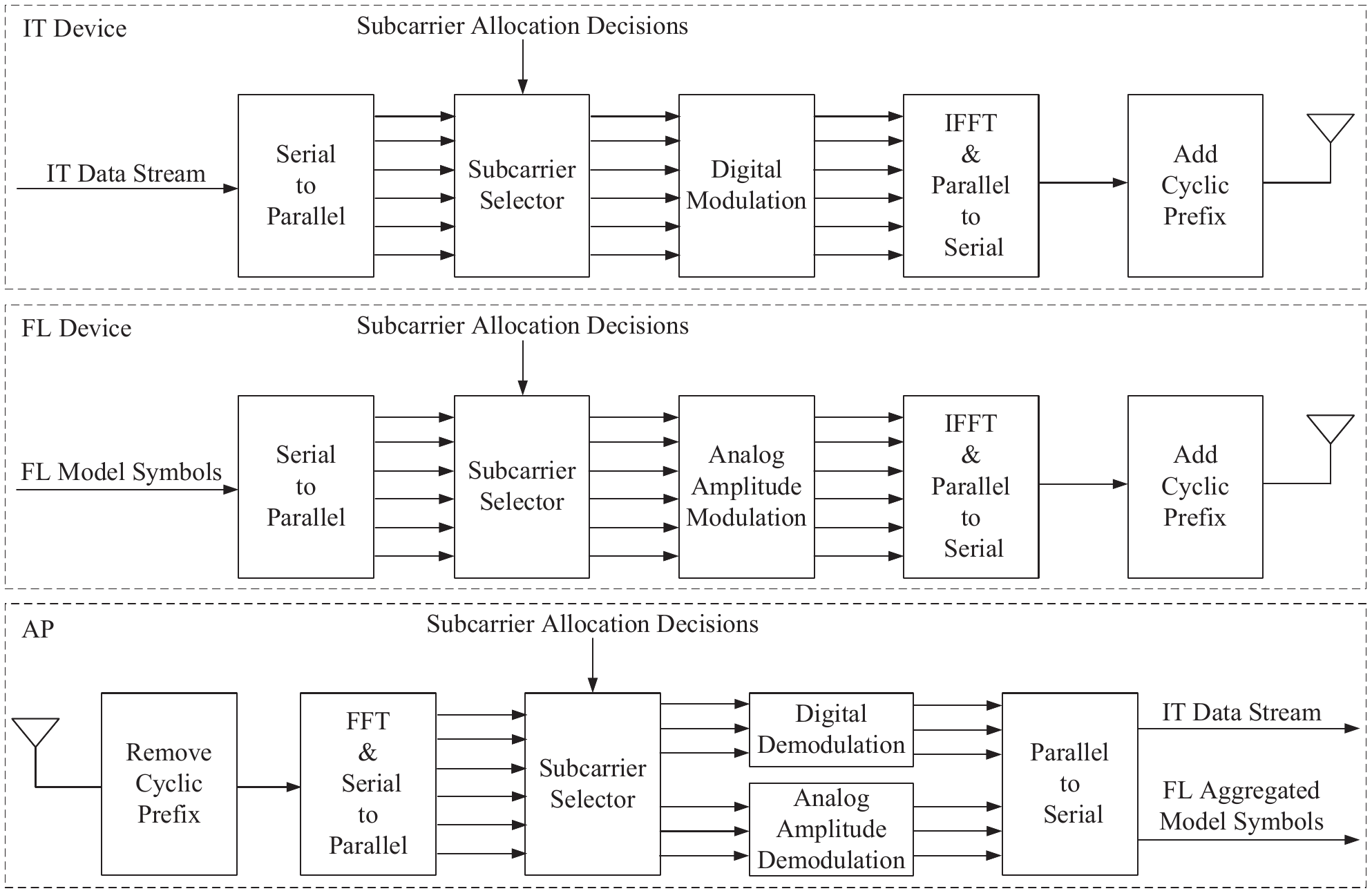}
		\caption{The workflow of the transmitters and receiver.} \label{fig:ofdm}
	\end{figure*}

	The workflow of the transmitters and the receiver in the CFLIT system is shown in Fig. \ref{fig:ofdm}. On one hand, the data stream of each IT device is first segmented into parallel streams, which are then modulated into digital signals on the allocated subcarriers before being transmitted to the AP. On the other hand, to leverage over-the-air computation for model aggregation, the local model symbols of FL devices are modulated into analog signals. We employ analog amplitude modulation \cite{zhu2019broadband} for the modulation, where each symbol is converted to a uniform waveform, e.g., a square wave, with amplitude proportional to the value of the original symbol. Due to the superposition nature of multiple-access channels, the received signal is approximately a square wave with amplitude proportional to the weighted sum of local symbols. Therefore, model aggregation can be efficiently achieved by transmitter-side pre-equalization. The receiver of the AP performs digital demodulation on the subcarriers allocated to IT and analog amplitude demodulation on the other subcarriers. As a result, the AP is able to extract data streams and over-the-air aggregated model symbols from the subcarriers accordingly.
	
	Define $o_{m, s}$ as the binary subcarrier allocation indicator for over-the-air FL on the $m$-th subcarrier of the $s$-th OFDM symbol. We have $o_{m, s} = 1$ if the $m$-th subcarrier of the $s$-th OFDM symbol is allocated to over-the-air model aggregation and $o_{m, s} = 0$ otherwise. Moreover, we define $b_{n, m, s}$ as the binary subcarrier allocation indicator for IT devices on the $m$-th subcarrier of the $s$-th OFDM symbol. That is, $b_{n, m, s} = 1$ if the $m$-th subcarrier of the $s$-th OFDM symbol is allocated to IT device $n$ and $b_{n, m, s} = 0$ otherwise. Consequently, we have $\sum_{n = 1}^N b_{n, m, s} + o_{m, s}  \leq 1$ for all $m, s$. In the following, we describe the over-the-air FL and IT models in detail.

	\subsection{Over-the-Air FL Model}
	We consider an over-the-air FL system, where the learning model is represented by a $d$-dimensional model parameter vector $\mathbf{w} \in \mathbb{R}^{d \times 1}$. Each FL device $k$ holds its local dataset $\mathcal{D}_k$ of size $D_k \triangleq |\mathcal{D}_k|$. Let $D \triangleq \sum_{k = 1}^K D_k$ denote the total number of training samples. We intend to minimize an empirical loss function of $\mathbf{w}$ given by
	\begin{equation}
		F(\mathbf{w}) = \sum_{k = 1}^K \rho_k F_k(\mathbf{w}; \mathcal{D}_k), \label{loss_function}
	\end{equation}
	where
	\begin{equation}
		F_k(\mathbf{w}; \mathcal{D}_k) = \frac{1}{D_k} \sum_{\mathbf{u} \in \mathcal{D}_k} f(\mathbf{w}; \mathbf{u})  \label{local_loss}
	\end{equation}
	is the local loss function with respect to the local dataset $\mathcal{D}_k$, $\rho_k \triangleq \frac{D_k}{D}$ is the corresponding weight of $F_k(\cdot)$, and $f(\mathbf{w}; \mathbf{u})$ is the loss function with respect to the local training sample $\mathbf{u}$.

	To minimize \eqref{loss_function}, the FL devices iteratively update the global model $\mathbf{w}$ via distributed stochastic gradient descent (SGD) with $T$ communication rounds. Specifically, the $t$-th communication round, $0 \leq t \leq T - 1$, consists of the following steps:
	\begin{enumerate}
		\item \emph{Model broadcasting:} The AP broadcasts the current global model $\mathbf{w}_{t}$ to all the FL devices in $\mathcal{K}$.
		
		\item \emph{Local model updating:} Each FL device $k$ parallelly updates the local model by minimizing \eqref{local_loss} via a $\tau$-step SGD. Specifically, the $l$-th step of local SGD, $1 \leq l \leq \tau$, updates the local model as
		\begin{equation}  \label{local_update}
				\mathbf{w}_{k, t}^{l + 1} = \mathbf{w}_{k, t}^{l} - \lambda_t \nabla F_k(\mathbf{w}_{k, t}^{l}, \xi_{k, t}^l),
		\end{equation}
		where $\mathbf{w}_{k, t}^1 = \mathbf{w}_t$, $\lambda_t$ is the learning rate in the $t$-th round, and $\nabla F_k(\mathbf{w}_{k, t}^{l}, \xi_{k, t}^l) \in \mathbb{R}^{d \times 1}$ is the gradient of $F_k(\cdot)$ with respect to the local mini batch $\xi_{k, t}^l \subseteq \mathcal{D}_k$ of size $B$. Here, each mini batch $\xi_{k, t}^l$ is uniformly drawn from $\mathcal{D}_k$ without replacement. After the $\tau$-step local update, each FL device $k$ computes its local model change $\bm{\Delta}_{k, t} \triangleq \mathbf{w}_{k, t}^{\tau + 1} - \mathbf{w}_{t}$.
		
		\item \emph{Over-the-air model aggregation:} All the FL devices concurrently upload the local model changes $\{\bm{\Delta}_{k, t}\}$ to the AP. The AP estimates the weighted sum $\mathbf{r}_t \triangleq \sum_{k = 1}^K \rho_k \bm{\Delta}_{k, t}$ by exploiting the signal-superposition property of wireless multiple-access channels and updates the global model $\mathbf{w}_{t + 1}$ as
		\begin{equation} \label{global_update}
				\mathbf{w}_{t + 1} = \mathbf{w}_{t} + \hat{\mathbf{r}}_t,
		\end{equation}
		where $\hat{\mathbf{r}}_t$ is the noisy estimate and $\hat{\mathbf{r}}_t \neq \mathbf{r}_t$ due to the inevitable communication noise.
	\end{enumerate}

	After $T$ communication rounds, we compute the weighted moving average of $\{\mathbf{w}_t: t = 0, \cdots, T - 1\}$ as \cite{stich2018local}
		\begin{equation}  \label{weighted_average}
			\hat{\mathbf{w}}_T \triangleq \frac{1}{S_T} \sum_{t = 0}^{T - 1} \eta_t \mathbf{w}_t,
		\end{equation}
	where $\eta_t$ is the weight of the $t$-th round, and $S_T \triangleq \sum_{t = 0}^{T - 1} \eta_t$.
	
	In this paper, we assume perfect downlink model broadcasting following \cite{cao2021optimized, zhang2021gradient, xu2021learning, liu2020privacy, amiri2020federated, shi2020joint} and focus on the uplink communication design for over-the-air model aggregation, which is elaborated as follows.
	
	With over-the-air model aggregation, we need $d$ RBs in each communication round to upload the $d$-dimensional vectors $\{\bm{\Delta}_{k, t}\}$ as each entry occupies one RB. Consequently, over-the-air FL requires $dT$ RBs in total to complete $T$ rounds of model aggregation.\footnote{Model compression by, e.g., gradient sparsification \cite{amiri2020machine, amiri2020federated, fan2021temporal}, can further reduce the dimensionality of local model changes. Note that our design is readily applicable to any over-the-air FL design that applies gradient sparsification by replacing the model change vectors with the sparsified gradient vectors. Accordingly, the total number of RBs for over-the-air model aggregation is reduced from $dT$ to $d^\prime T$, where $d^\prime < d$ represents the dimension of compressed local gradient vectors.} We define $\mathbf{\Phi} \triangleq \{(m, s) : o_{m, s} = 1, \forall m \in \mathcal{M}, s \in \mathcal{S}\}$ as the set of indices of all the subcarriers allocated to over-the-air FL, where $|\mathbf{\Phi}| = dT$. We partition $\mathbf{\Phi}$ into $T$ disjoint $d$-dimensional vectors, where the $t$-th vector $\mathbf{\Phi}_t$ denotes the RB indices assigned to the $t$-th model aggregation round. For example, the first $d$ RBs in $\mathbf{\Phi}$ is grouped as $\mathbf{\Phi}_1$. We denote the entries of $\mathbf{\Phi}_t$ by $\mathbf{\Phi}_{t} \triangleq [\Phi[dt + 1], \cdots, \Phi[d(t + 1)] ]^T$.
	
	In the $t$-th model aggregation round, to facilitate the pre-equalization design and transmit power control, the local model change vector of device $k$, $\bm{\Delta}_{k, t} \in \mathbb{R}^{d \times 1}$, is transformed into a normalized transmit symbol vector $\mathbf{x}_{k, t} \in \mathbb{R}^{d \times 1}$ such that $\mathbb{E}[\mathbf{x}_{k, t} \mathbf{x}_{k', t}^H] = \mathbf{0}$ and $\mathbb{E}[\mathbf{x}_{k, t} \mathbf{x}_{k, t}^H] = \mathbf{I}_d, \forall k, k' \in \mathcal{K}, k \neq k'$ \cite{yang2020federated, zhu2019broadband}. Specifically, we first compute the local mean and variance of the local model change $\bm{\Delta}_{k, t}$ at each FL device $k$ by
		\begin{equation}
			\overline{\Delta}_{k, t} = \frac{1}{d} \sum_{i = 1}^{d} \Delta_{k, t}[i],  ~~~ \nu_{k, t}^2 = \frac{1}{d} \sum_{i = 1}^{d} \left(\Delta_{k, t}[i] - \overline{\Delta}_{k, t} \right)^2,
		\end{equation}
	where $\Delta_{k, t}[i]$ is the $i$-th entry of $\bm{\Delta}_{k, t}$. Then, each FL device $k$ uploads its local statistics $\overline{\Delta}_{k, t}$ and $\nu_{k, t}^2$ to the AP\footnote{The local statistics $\{\overline{\Delta}_{k, t}, \nu_{k, t}^2\}$ can be uploaded to the AP via conventional orthogonal transmission with an overhead proportional to the number of FL devices $K$. Since a learning model typically contains thousands or even millions of parameters (i.e., $d \gg K$), we ignore the overhead on
	uploading the local statistics.} and computes
		\begin{equation}
			x_{k, t}[i] = \frac{\Delta_{k, t}[i] - \overline{\Delta}_{k, t}}{\nu_{k, t}},
		\end{equation}
	where $x_{k, t}[i]$ is the $i$-th entry of $\mathbf{x}_{k, t}$. Let $h_{k, m, s} \in \mathbb{C}$ denote the channel coefficient between FL device $k$ and the AP on the $m$-th subcarrier of the $s$-th OFDM symbol, and $p_{k, m, s} \in \mathbb{C}$ denote the transmit scalar at FL device $k$ on the $m$-th subcarrier of the $s$-th OFDM symbol. The channel coefficient and transmit scalar for transmitting $x_{k, t}[i]$ over the RB indexed by $\Phi_t[i]$ are denoted by $h_{k, \Phi_{t}[i]}$ and $p_{k, \Phi_{t}[i]}$, respectively. Let $P_1$ denote the maximum transmit power of each FL device on each subcarrier. The transmit power at each FL device is constrained by
		\begin{equation}
			\mathbb{E} \left[|p_{k, \Phi_{t}[i]} x_{k, t}[i]|^2 \right]= |p_{k, \Phi_{t}[i]}|^2 \leq P_1, ~~\forall k \in \mathcal{K}, \Phi_{t}[i].  \label{ak_cons}
		\end{equation}
	We denote the received signal vector at the AP in the $t$-th round by $\mathbf{y}_t = [y_t[1], \cdots, y_t[d]]^T$, with the $i$-th entry being
		\begin{equation}  \label{received_signal}
			y_t[i] = \sum_{k = 1}^{K} h_{k, \Phi_{t}[i]} p_{k, \Phi_{t}[i]} x_{k, t}[i] + z_t[i],
		\end{equation}
	where $z_t[i] \sim \mathcal{CN}(0, \sigma^2)$ is the additive white Gaussian noise (AWGN) with variance $\sigma^2$.
	The AP estimates the weighted sum $\sum_{k = 1}^{K} \rho_k \Delta_{k, t} [i]$ from $y_t[i]$ by a linear estimator as
		\begin{align} \label{aggregated_signal}
			\hat{r}_t[i] =& c_t[i] y_t[i] + \overline{\Delta}_{t}  \nonumber \\
			=& c_t[i] \sum_{k = 1}^K \frac{h_{k, \Phi_{t}[i]} p_{k, \Phi_{t}[i]}}{\nu_{k, t}} \left(\Delta_{k, t}[i] - \overline{\Delta}_{k, t} \right) + c_t[i] z_t[i]  \nonumber \\
			&+ \overline{\Delta}_{t},
		\end{align}
	where $c_t[i] \in \mathbb{C}$ is the de-noising multiplicative scalar and $\overline{\Delta}_{t} \triangleq \sum_{k = 1}^{K} \rho_k \overline{\Delta}_{k, t}$ is used for de-normalization. With \eqref{aggregated_signal}, the AP collects $\hat{\mathbf{r}}_t = [\hat{r}_t[1], \cdots, \hat{r}_t[d]]$ and updates the global model by \eqref{global_update}.

	\subsection{IT Model}
	As over-the-air FL requires $dT$ RBs in total, the remaining $MS - dT$ RBs are assigned to IT devices for uplink data transmissions, where each RB is allocated to one IT device. Let $g_{n, m, s} \in \mathbb{C}$ denote the channel coefficient between IT device $n$ and the AP on the $m$-th subcarrier of the $s$-th OFDM symbol. Then, the corresponding received signal on the $m$-th subcarrier of the $s$-th OFDM symbol can be expressed as
		\begin{equation}
			y_{m, s} = \sum_{n = 1}^N b_{n, m, s} g_{n, m, s} \sqrt{P_2} x_{n, m, s} + z_{m, s},
		\end{equation}
	where $x_{n, m, s}$ is the digitally-modulated transmit signal with $\mathbb{E}[|x_{n, m, s}|^2] = 1$, $P_2$ is the transmit power of each IT device, and $z_{m, s} \sim \mathcal{CN}(0, \sigma^2)$ is the AWGN. As a result, the achievable rate of IT device $n$ on the $m$-th subcarrier of the $s$-th OFDM symbol is given by
		\begin{equation} \label{achievable_rate}
			R_{n, m, s} = b_{n, m, s} \log_2 \left( 1 + \frac{P_{2} |g_{n, m, s}|^2}{\phi \sigma^2} \right),
		\end{equation}
	where $\phi \geq 1$ is the achievable rate gap due to channel coding and modulation. Consequently, the average IT sum-rate on each OFDM RB is given by
		\begin{align}  \label{sum_rate}
			R_{\text{avg}} =& \frac{1}{M S} \sum_{s = 1}^S \sum_{m = 1}^M \sum_{n = 1}^N R_{n, m, s}  \nonumber \\
			=& \frac{1}{M S} \sum_{s = 1}^S \sum_{m = 1}^M \sum_{n = 1}^N b_{n, m, s} \log_2 \bigg( 1 + \frac{P_{2} |g_{n, m, s}|^2}{\phi \sigma^2} \bigg).
		\end{align}

	\subsection{Problem Formulation}
	We consider the joint optimization of radio resource allocation and the hyper-parameters of FL, including the number of communication rounds $T$ and the number of local updates $\tau$. To facilitate the problem formulation, we first introduce the definition of an $\epsilon$-accurate solution to characterize the ultimate target of FL model training.
	
	\emph{\textbf{Definition 1} ($\epsilon$-accurate solution).} The output of the FL training process $\hat{\mathbf{w}}_T$ is an $\epsilon$-accurate solution to the FL loss minimization problem $\min_{\mathbf{w}} F(\mathbf{w})$ if
		\begin{equation}  \label{epsilon_sol}
			F(\hat{\mathbf{w}}_T) - F^* \leq \epsilon,
		\end{equation}
	where $F^* \! \triangleq \! F(\mathbf{w}^* \! )$, $\mathbf{w}^*$ is the global optimum of $F(\cdot)$, and $\epsilon \! > \! 0$ is a given positive small number.
	
	The value of $\epsilon$ in \eqref{epsilon_sol} measures the tolerance of FL model training. A smaller $\epsilon$ leads to a smaller optimality gap and thus a better training performance. We note that the training performance $F(\hat{\mathbf{w}}_T)$ is determined by the model aggregation quality as well as the learning design, including the choices of the communication rounds $T$ and the number of local updates $\tau$. In particular, as discussed in the Introduction, the value of $\tau$ determines the computation-to-communication ratio and critically impacts $T$ to achieve a certain training performance. Furthermore, a larger $T$ implies more RBs needed for over-the-air FL training, leaving fewer RBs for data transmissions of IT devices.
	
	Besides the learning hyper-parameters $\tau$ and $T$, the FL training performance and the IT data rate are both affected by the wireless channel fading and thus depend on the subcarrier allocation decisions. On one hand, wireless channel fading affects the channel gains and thus the data rate of the IT process as given in \eqref{sum_rate}. On the other hand, the transceiver design variables $\{p_{k, \Phi_{t}[i]}, c_t[i]\}$ of over-the-air FL jointly affect the quality of the model aggregation and the resulting global model update. However, we emphasize that FL and IT have different sensitivities against communication errors. The gradient descent step in FL is naturally robust against a small gradient error as slightly perturbing the descent direction does not affect the convergence to the local minimum; see \cite[Fig. 5]{soudry2015memristor}. To better utilize the radio resources to support the coexistence of FL and IT, we exploit the different error sensitivities of FL and IT and allocate the subcarriers to IT if the IT devices have good channel conditions. The remaining radio resources are then utilized for FL model aggregation. To this end, we propose to \emph{maximize the IT data rate} and at the same time \emph{constrain the expected output of the FL training process to be $\epsilon$-accurate} in $S$ symbol durations. Note that the FL subcarrier allocation affects the FL communication rounds $T$, and hence is coupled with the IT subcarrier allocation. By taking the communication variables $\{b_{n, m, s}, o_{m, s}, p_{k, \Phi_{t}[i]}, c_t[i] \}$ and the learning hyper-parameters $\{T, \tau \}$ into account, we formulate the CFLIT system optimization problem as:
		\begin{subequations} \label{sflit_problem}
			\begin{align}
				\hspace{-0.5em}\text{(P1):} \ \max_{ \underset{\{p_{k, \Phi_{t}[i]}, c_t[i] \}}{\{b_{n, m, s}, o_{m, s}, T, \tau \}}}~&  R_{\text{avg}} \label{obj} \\
				{\rm s.t.}~~~
				&\mathbb{E} \left[F(\hat{\mathbf{w}}_T; \tau) - F^* \right] \leq \epsilon,  \label{cons_1}  \\							
				&\sum_{s = 1}^S \sum_{m = 1}^M o_{m, s} \geq dT,  \label{cons_2}  \\
				&\sum_{n = 1}^N b_{n, m, s} + o_{m, s}  \leq 1, \forall m, s, \label{cons_3}  \\
				&b_{n, m, s} \in \{0, 1\}, \forall n, m, s,  \label{cons_4}  \\
				&o_{m, s} \in \{0, 1\}, \forall m, s,  \label{cons_5}  \\	
				&|p_{k, \Phi_{t}[i]}|^2 \leq P_1, \forall k \in \mathcal{K}, \Phi_{t}[i],  \label{cons_6}  \\	
				&T, \tau \in \mathbb{Z}^+,  \label{cons_7}
			\end{align}
		\end{subequations}
	where the expectation is taken with respect to the local mini-batch samples, the communication noise, and the channel fading coefficients. $F(\hat{\mathbf{w}}_T; \tau)$ is the achieved training loss after $T$ FL communication rounds with the number of local updates given by $\tau$. In Problem (P1), the constraint \eqref{cons_1} guarantees an $\epsilon$-accurate solution of FL training, \eqref{cons_2} ensures a sufficient number of allocated RBs for over-the-air FL, \eqref{cons_3} implies that each OFDM RB is assigned to either the over-the-air model aggregation or at most one IT data stream, and \eqref{cons_6} represents the transmit power constraints at FL devices.
	
	The difficulty of solving Problem (P1) is twofold. First, to satisfy the FL convergence requirement in \eqref{cons_1}, we must understand how the learning hyper-parameters $(T, \tau)$, the transceiver design variables $\{p_{k, \Phi_{t}[i]}, c_t[i]\}$, and the subcarrier allocation decisions $\{o_{m, s}\}$ jointly affect the achieved training loss $F(\hat{\mathbf{w}}_T; \tau)$ of over-the-air FL. However, there does not exist an explicit analytical expression to characterize the above coupled impact in the existing literature. This makes it difficult to quantify the required number of RBs for over-the-air model aggregation in advance and minimize it accordingly to optimize the IT performance. Second, to solve (P1) optimally, we need to know the CSI for the entire horizon $s = 1, \cdots, S$ at the beginning of time. In practice, however, such non-causal CSI is not available. That is, we need to optimize $\{b_{n, m, s}, o_{m, s}\}$ for each symbol duration $s$ based on the instantaneous CSI that is observed at that time. Therefore, an online RB allocation scheme that only depends on the current channel coefficients is required. In the next section, we analyze the performance of over-the-air FL with an arbitrary choice of $(T, \tau)$ under mild assumptions on the learning model. To minimize the radio resource consumption of over-the-air FL, we optimize the transceiver design variables $\{p_{k, \Phi_{t}[i]}, c_t[i]\}$ and use the analytical result to compute the optimal solution of $(T, \tau)$ to (P1) in the asymptotic regime. Then, in Section IV, we propose a low-complexity online algorithm for subcarrier allocation and analyze its performance.

	\section{Convergence Analysis and Learning Optimization}	
	In this section, we analyze the convergence of over-the-air FL. To facilitate the analysis, we first present the assumptions on the local loss functions and analyze the model aggregation error in over-the-air model aggregation. Based on this, we show a sufficient condition for FL convergence and derive an asymptotic upper bound of the learning performance in terms of $\mathbb{E} [F(\hat{\mathbf{w}}_T) - F^*]$. Finally, we use the derived bound to optimize the learning hyper-parameters $\tau$ and $T$.
	
	\subsection{Assumptions and Preliminaries}	
	We make the following assumptions on the local loss functions $F_1, \cdots, F_K$:
	
	\textbf{Assumption 1:} \emph{$F_1, \cdots, F_K$ are all $\mu$-strongly convex, i.e., $\forall \mathbf{w}, \mathbf{w}^\prime \in \mathbb{R}^{d \times 1},$}
		\begin{equation}  \label{strong_convexity}
			F_k(\mathbf{w}) \geq F_k(\mathbf{w}^\prime) + (\mathbf{w} - \mathbf{w}^\prime)^T \nabla F_k(\mathbf{w}^\prime) + \frac{\mu}{2} \|\mathbf{w} - \mathbf{w}^\prime\|_2^2.
		\end{equation}
	
	\textbf{Assumption 2:} \emph{$F_1, \cdots, F_K$ are all $L$-smooth, i.e., $\forall \mathbf{w}$, $\mathbf{w}^\prime \in \mathbb{R}^{d \times 1},$}
		\begin{equation}  \label{L_smoothness}
			F_k(\mathbf{w}) \leq F_k(\mathbf{w}^\prime) + (\mathbf{w} - \mathbf{w}^\prime)^T \nabla F_k(\mathbf{w}^\prime) + \frac{L}{2} \|\mathbf{w} - \mathbf{w}^\prime\|_2^2.
		\end{equation}
	
	\textbf{Assumption 3:} \emph{The expected $l_2$-norm of the stochastic gradient $\nabla F_k(\mathbf{w}_{k, t}^l, \xi_{k, t}^l)$ is upper bounded, i.e., $\forall k, t, l,$}
		\begin{equation}  \label{square_norm}
			\mathbb{E} [\|\nabla F_k(\mathbf{w}_{k, t}^l, \xi_{k, t}^l)\|_2^2 ] \leq G^2,
		\end{equation}
	where the expectation is taken with respect to the local mini-batch samples, the communication noise, and the channel fading coefficients.
	
	\begin{remark}
		The above assumptions are commonly adopted in the literature; see, e.g., \cite{stich2018local, li2019convergence, amiri2021convergence}. In practice, Assumption 3 can be guaranteed by imposing an additional gradient clipping operation \cite{pascanu2013difficulty}.
	\end{remark}
	
	The following definition defines a metric of data heterogeneity.
	
	\emph{\textbf{Definition 2} ($\Gamma$-data heterogeneity).} The degree of data heterogeneity across the FL devices is quantified by
		\begin{equation}  \label{Gamma}
			\Gamma \triangleq F^* - \sum_{k = 1}^K \rho_k F_k^*,
		\end{equation}
	where $F_k^*$ denotes the minimum value of local loss function $F_k(\cdot)$ at FL device $k$, and $\Gamma \geq 0$.
	
	As the number of local training samples goes to infinity, i.e., $D_k \rightarrow \infty, \forall k \in \mathcal{K}$, the locally trained models converge to the global model when the training data are i.i.d. distributed, i.e., $\mathbf{w}_k^* \rightarrow \mathbf{w}^*, \forall k \in \mathcal{K}$, and thus $\Gamma \rightarrow 0$. For non-i.i.d. data distribution, $\Gamma \neq 0$. The larger $\Gamma$, the more heterogeneous the data distribution is.
	
	As shown in Section II-B, over-the-air model aggregation constructs a noisy estimate $\hat{\mathbf{r}}_t$ at the AP using \eqref{aggregated_signal}. The estimation accuracy of $\hat{\mathbf{r}}_t$ affects the convergence of over-the-air FL because the estimation error leads to inaccurate global model update. We quantify the estimation accuracy by the MSE between $\hat{\mathbf{r}}_t$ and the ground-truth $\mathbf{r}_t$ as
		\begin{equation}
			e_{t} = \mathbb{E}[\|\hat{\mathbf{r}}_t - \mathbf{r}_t \|_2^2],
		\end{equation}
	where the expectation is taken with respect to the communication noise.
	
	It is shown in \cite{liu2020reconfigurable} that a larger model aggregation error $e_t$ leads to a smaller convergence rate of over-the-air FL. Therefore, we can optimize the transceiver design variables $\{p_{k, \Phi_{t}[i]}, c_t[i]\}$ by minimizing $e_t$, as detailed by the following lemma.
		\begin{lemma} \label{MMSE}
				Given the channel coefficients $\{h_{k, \Phi_{t}[i]}\}$, the optimal $p_{k, \Phi_{t}[i]}$ and $c_t[i]$ are
					\begin{equation} \label{opt_p_c}
						p_{k, \Phi_{t}[i]} = \frac{\rho_k \nu_{k, t}}{c_t[i] h_{k, \Phi_{t}[i]}}, \; \forall k \in \mathcal{K},   c_t[i] = \frac{1}{\sqrt{P_1}} \max_{k \in \mathcal{K}} \frac{\rho_k \nu_{k, t}}{|h_{k, \Phi_{t}[i]}|}.
					\end{equation}
				Moreover, the minimum $e_{t}$ is given by
					\begin{equation} \label{MMSE_eq}
						e_{t} = \frac{\sigma^2}{P_1}  \sum_{i = 1}^d \max_{k \in \mathcal{K}} \frac{\rho_k^2 \nu_{k, t}^2}{|h_{k, \Phi_{t}[i]}|^2}.
					\end{equation}
			\end{lemma}
		\begin{IEEEproof}
			See Appendix \ref{MMSE_proof}.
		\end{IEEEproof}
	
	Lemma \ref{MMSE} gives the optimal solution to $\{p_{k, \Phi_{t}[i]}, c_t[i]\}$ and the analytical expression of the minimum model aggregation error $e_{t}$ for given channel coefficients $\{h_{k, \Phi_{t}[i]}\}$. To facilitate the convergence analysis, we need a tractable expression for the model aggregation error $e_t$ that is independent of the instantaneous channel coefficients. Based on Lemma \ref{MMSE} and Assumption 3, we obtain an upper bound of the expected value of $e_t$ in the following lemma.
	\begin{lemma} \label{MMSE2}
		With Assumption 3, we have
		\begin{equation}  \label{MMSE2_eq}
				\mathbb{E} [ e_t ] \leq \lambda_t^2 \frac{\tau^2 G^2 \sigma^2}{P_1} \mathbb{E}\left[ \max_{k \in \mathcal{K}} \frac{\rho_k^2}{|h_{k}|^2} \right],
		\end{equation}
		where the expectations are taken with respect to the local mini-batch samples and the channel fading coefficients, and $\{h_k, \forall k \in \mathcal{K}\}$ are \emph{virtual} i.i.d. random variables following the Gaussian distribution of $\mathcal{CN}(0, 1)$.
	\end{lemma}
	\begin{IEEEproof}
		See Appendix \ref{MMSE2_proof}.
	\end{IEEEproof}

	\subsection{Convergence Result}
	With Lemma \ref{MMSE2} and Assumptions 1-3, we derive an upper bound of $\mathbb{E} [F(\hat{\mathbf{w}}_T) - F^*]$ in the following theorem.
	
	\begin{theorem}  \label{theorem1}
		Let the learning rate $\lambda_t = \frac{8}{\mu \tau (\gamma + t)}$ and the corresponding weight of the $t$-th global model $\eta_t = (\gamma + t)^2$ with $\gamma \geq \frac{16 L}{\mu}$. With Assumptions 1-3, we have
			\begin{align} \label{convergence}
				\mathbb{E} [&F(\hat{\mathbf{w}}_T) - F^*] \leq \frac{8T(T + 2 \gamma)}{\mu S_T} \bigg( \frac{2\tau^2 + 1}{3 \tau} G^2 + \frac{4 L \Gamma}{\tau}  \nonumber \\
				&+ \frac{G^2 \sigma^2}{P_1} \mathbb{E} \left[ \max_{k \in \mathcal{K}} \frac{\rho_k^2}{|h_k|^2} \right] \bigg) + \frac{\mu \gamma^3}{4 S_T} \mathbb{E} [\| \mathbf{w}_{0} - \mathbf{w}^* \|_2^2],
			\end{align}
		where $S_T$ is defined in \eqref{weighted_average}, and the expectations are taken with respect to the local mini-batch samples, the communication noise, and the channel fading coefficients.
	\end{theorem}
	\begin{IEEEproof}
		See Appendix \ref{theorem1_proof}.
	\end{IEEEproof}
	
	From Theorem \ref{theorem1}, we see that $\mathbb{E} [F(\hat{\mathbf{w}}_T) - F^*]$ has an upper bound related to the term $\mathbb{E} \big[ \max_{k \in \mathcal{K}} \frac{\rho_k^2}{|h_k|^2} \big]$, which refers to the expected model aggregation error. Since the expected error is independent of the instantaneous channel coefficients $\{h_{k, \Phi_{t}[i]}\}$, the result in Theorem \ref{theorem1} applies to any subcarrier allocation scheme. However, Theorem \ref{theorem1} does not explicitly reveal the effect of $\tau$ and $T$. To this end, we further characterize the convergence behavior of $\mathbb{E} [F(\hat{\mathbf{w}}_T) - F^*]$ in the large-system limit, i.e., $T \rightarrow \infty$, in the following corollary.
	\begin{corollary}  \label{corollary1}
		Suppose that the assumptions and conditions in Theorem \ref{theorem1} hold. We have
		\begin{equation} \label{corollay_eq1}
			\mathbb{E} [F(\hat{\mathbf{w}}_T) - F^*] \leq \frac{\zeta(\tau)}{T} + \frac{2 \gamma \zeta(\tau)}{T^2} + \frac{3 \mu \gamma^3  \mathbb{E} [\| \mathbf{w}_{0} - \mathbf{w}^* \|_2^2]}{4 T^3},
		\end{equation}
		where
		\begin{equation} \label{corollay_eq2}
			\zeta(\tau) \triangleq \frac{24}{\mu} \left(\frac{2\tau^2 + 1}{3 \tau} G^2 + \frac{4 L \Gamma}{\tau} + \frac{G^2 \sigma^2}{P_1} \mathbb{E} \left[ \max_{k \in \mathcal{K}} \frac{\rho_k^2}{|h_k|^2} \right] \right).
		\end{equation}
		Moreover, when $T$ is sufficiently large, i.e., $T \rightarrow \infty$, the last two terms on the right hand side (r.h.s.) of \eqref{corollay_eq1} are diminishing as $\frac{1}{T^2}, \frac{1}{T^3} \rightarrow 0$. In this case, we can approximate the upper bound of $\mathbb{E} [F(\hat{\mathbf{w}}_T) - F^*]$ as
		\begin{equation} \label{corollay_eq3}
				\mathbb{E} [F(\hat{\mathbf{w}}_T) - F^*] \leq \frac{\zeta(\tau)}{T} + O \left(\frac{1}{T^2} \right) \approx \frac{\zeta(\tau)}{T}  \quad \text{as} \quad T \rightarrow \infty.
		\end{equation}		
	\end{corollary}
	\begin{IEEEproof}
		Since $S_T = \sum_{t = 0}^{T - 1} \eta_t = \sum_{t = 0}^{T - 1} (\gamma + t)^2$ and $\gamma \geq \frac{16 L}{\mu} \geq 1$, we have
		\begin{align}
			S_T =& \sum_{t = 0}^{T - 1} \gamma^2 + 2 \gamma t + t^2  \nonumber \\
			=& \gamma^2T + \gamma T(T - 1) + \frac{T(T - 1) (2T - 1)}{6}  \nonumber \\
			\geq& \frac{1}{3} T^3.
		\end{align}
	Plugging this result into \eqref{convergence}, we obtain \eqref{corollay_eq1}.
	\end{IEEEproof}

	From Corollary \ref{corollary1}, we see that the output of the FL training process $\hat{\mathbf{w}}_T$ guarantees to converge to the global optimum and over-the-air FL has a convergence rate of $O(\frac{1}{T})$ for a sufficiently large $T$. Moreover, as shown in \eqref{corollay_eq3}, the asymptotic upper bound in the large-system limit is a function of the learning hyper-parameters $\tau$ and $T$. Since $T$ is sufficiently large in practice, we use the upper bound in \eqref{corollay_eq3} as an approximation of $\mathbb{E} [F(\hat{\mathbf{w}}_T) - F^*]$, i.e., $\mathbb{E} [F(\hat{\mathbf{w}}_T) - F^*] \approx \frac{\zeta(\tau)}{T}$, and asymptotically optimize $\tau$ and $T$ in the next subsection.
	
	\subsection{Optimization of Learning Parameters}
	To obtain an $\epsilon$-accurate solution, i.e., to achieve $\mathbb{E} [F(\hat{\mathbf{w}}_T) - F^*] \leq \epsilon$, it is sufficient to ensure $\frac{\zeta(\tau)}{T} \leq \epsilon$, which yields
		\begin{align} \label{T_bound}
			T \geq& \frac{\zeta(\tau)}{\epsilon}  \nonumber \\
			=& \frac{24}{\mu \epsilon} \left(\frac{2 G^2}{3} \tau + \frac{G^2 + 12 L \Gamma}{3 \tau} + \frac{G^2 \sigma^2}{P_1} \mathbb{E} \left[ \max_{k \in \mathcal{K}} \frac{\rho_k^2}{|h_k|^2} \right] \right).
		\end{align}
	The last term on the r.h.s. of \eqref{T_bound} reflects the effect of the over-the-air model aggregation: A smaller error in model aggregation (i.e., $e_t$) leads to a smaller number of required training rounds $T$. From \eqref{T_bound}, we see that $T$ is lower bounded by a function of $\tau$, which is not monotonic. Specifically, the r.h.s. of \eqref{T_bound} first decreases and then increases with $\tau$, showing that the minimizer of $\zeta(\tau)$ exists. Note that we can equivalently minimize the radio resource consumption of over-the-air FL by minimizing the required communication rounds $T$. To obtain the minimum $T^*$, it suffices to find the optimal $\tau^*$ that minimizes $\zeta(\tau)$, i.e., $\tau^* = \mathop{\arg\min} \zeta (\tau)$, which is equivalent to solving the following optimization problem:
		\begin{subequations} \label{opt_tau_int}
			\begin{align}
				\min_{\tau} \quad &\psi(\tau) = \frac{2 G^2}{3} \tau  + \frac{G^2 + 12 L \Gamma}{3 \tau}  \label{opt_tau_obj} \\
				{\rm s.t.}~~~	
				&\tau \in \mathbb{Z}^+, \tau \geq 1,  \label{opt_tau_cons1}
			\end{align}
		\end{subequations}
	where the objective function $\psi(\tau)$ is derived from $\zeta(\tau)$ by omitting the constant term.
	
	Problem \eqref{opt_tau_int} is an one-dimensional integer programming problem and can be solved optimally as follows. Specifically, we first solve the following relaxed problem:
		\begin{equation} \label{opt_tau_relax}
			\min_{\tau \in \mathbb{R}, \tau \geq 1} \quad \psi(\tau) = \frac{2 G^2}{3} \tau  + \frac{G^2 + 12 L \Gamma}{3 \tau}.
		\end{equation}
	Since Problem \eqref{opt_tau_relax} is convex, the optimal solution is given by
		\begin{equation}  \label{tau_relax}
				\tau_{\textrm{relax}} = \max \left\{1, \sqrt{\frac{1}{2} + \frac{6 L \Gamma}{G^2}} \right\}.
		\end{equation}
	Then, the optimal solution $\tau^*$ to Problem \eqref{opt_tau_int} can be obtained by simply rounding
	\eqref{tau_relax} as
		\begin{equation}  \label{opt_tau}
			\tau^* = \argmin_{\tau \in \{\lfloor \tau_{\textrm{relax}} \rfloor, \lceil \tau_{\textrm{relax}} \rceil \}} \psi(\tau),
		\end{equation}
	where $\lfloor \cdot \rfloor$ denotes the floor function, and $\lceil \cdot \rceil$ denotes the ceiling function.
	
	Substituting the optimal value of $\tau^*$ in \eqref{opt_tau} into \eqref{T_bound}, we have
		\begin{align}  \label{T_relax}
			T \geq& \frac{\zeta(\tau^*)}{\epsilon}  \nonumber \\
			=& \frac{24}{\mu \epsilon} \left(\frac{2 G^2}{3} \tau^* + \frac{G^2 + 12 L \Gamma}{3 \tau^*} + \frac{G^2 \sigma^2}{P_1} \mathbb{E} \left[ \max_{k \in \mathcal{K}} \frac{\rho_k^2}{|h_k|^2} \right] \right).
		\end{align}
	As $T \in \mathbb{Z}^+$, the minimum $T^*$ satisfying \eqref{T_relax} is given by
		\begin{equation}  \label{opt_T}
			T^* \!=\! \left\lceil \! \frac{24}{\mu \epsilon} \left(\! \frac{2 G^2}{3} \tau^* + \frac{G^2 + 12 L \Gamma}{3 \tau^*} + \frac{G^2 \sigma^2}{P_1} \mathbb{E} \left[ \max_{k \in \mathcal{K}} \frac{\rho_k^2}{|h_k|^2} \right] \!\right) \!\right\rceil .
		\end{equation}
	
		We see from \eqref{tau_relax} and \eqref{opt_T} that the large-system asymptotic results $\tau^*$ and $T^*$ depend on the parameters $L, \Gamma$, and $G$, which are determined by the data distribution and the learning task. The accuracy of the derived $\{\tau^*, T^*\}$ is verified by numerical results in Section V.
	
	\section{Online Subcarrier Allocation for CFLIT}
	We have discussed the optimization of $\{T, \tau, p_{k, \Phi_{t}[i]}, c_t[i]\}$ in Section III. The remaining challenge of solving Problem (P1) lies in optimizing the binary subcarrier allocation variables $\{b_{n, m, s}, o_{m, s} \}$ without the non-causal CSI. In this section, we propose a low-complexity threshold-based online algorithm that only requires the CSI in the current symbol duration $s$ to optimize $\{b_{n, m, s}, o_{m, s} \}$ in an online manner. Furthermore, we analyze the achievable performance of the proposed algorithm.

	\subsection{Threshold-Based Online Subcarrier Allocation}
	As discussed in Section III, our analysis of FL convergence does not rely on the instantaneous CSI of FL devices. Moreover, the channel coefficients of FL devices are independent of those of IT devices. Therefore, the convergence analysis in Section III is applicable to arbitrary subcarrier allocation. With the optimized $\{T, \tau, p_{k, \Phi_{t}[i]}, c_t[i]\}$, the FL convergence requirement in \eqref{cons_1} can be satisfied by ensuring that the number of allocated subcarriers for over-the-air FL is sufficiently large to complete $T^*$ rounds of model aggregation. Accordingly, Problem (P1) is simplified as:
	\begin{subequations} \label{sflit_problem2}
				\begin{align}
				\hspace{-6em}\text{(P2):} \max_{\{b_{n, m, s}, o_{m, s} \}}~& R_{\text{avg}}  \label{P2_obj} \\
				{\rm s.t.}\qquad&\sum_{s = 1}^S \sum_{m = 1}^M o_{m, s} \geq dT^*,  \label{cons2_2}  \\
				&\sum_{n = 1}^N b_{n, m, s} + o_{m, s}  \leq 1, \forall m, s, \label{cons2_3}  \\
				&b_{n, m, s} \in \{0, 1\}, \forall n, m, s,  \label{cons2_4}  \\
				&o_{m, s} \in \{0, 1\}, \forall m, s. \label{cons2_5}
		\end{align}
	\end{subequations}

	We note that the subcarrier allocation is independent of the channel coefficients of FL devices $\{h_{k, m, s}\}$, as implied by \eqref{P2_obj} and \eqref{cons2_2}. When the CSI of the entire time horizon is available, a simple offline solution to (P2) follows. Denote the highest channel gain among all the IT devices on the $m$-th subcarrier of the $s$-th OFDM symbol by $\bar{g}_{m, s}$ as
		\begin{equation} \label{max_g}
			\bar{g}_{m, s} = \max_{n \in \mathcal{N}} \left\{|g_{1, m, s}|^2, \cdots, |g_{n, m, s}|^2, \cdots, |g_{N, m, s}|^2 \right\}.
		\end{equation}
	The optimal subcarrier allocation decision $\{b_{n, m, s}, o_{m,s}\}$ is given by sorting $MS$ channel gains $\bar g_{m,s}$ in the descending order and assigning the top $MS-dT^*$ subcarriers to IT and the last $dT^*$ to over-the-air FL. Meanwhile, for any $o_{m,s}=0$, the subcarrier is assigned to the IT device with the highest channel gain. In practice, however, such CSI is unavailable in advance. This makes it difficult to exactly find these globally best subcarriers in each symbol duration. In the following, we propose a threshold-based online subcarrier allocation algorithm that only requires the current CSI to mimic the offline optimal solution.
	
	Since the channel coefficient $g_{n, m, s}$ follows the i.i.d. complex Gaussian distribution of $\mathcal{CN}(0, 1)$, the channel gain $|g_{n, m, s}|^2, \forall n, m, s,$ follows the i.i.d. chi-squared distribution with 2 degrees of freedom, i.e., $|g_{n, m, s}|^2 \sim \chi^2(2)$, whose cumulative distribution function (CDF) is specified by
		\begin{align}  \label{g_cdf}
			F_{|g_{n, m, s}|^2}(x) = P(|g_{n, m, s}|^2 \leq x) = 1 - e^{-x}.
		\end{align}
	With \eqref{g_cdf}, the CDF of $\bar{g}_{m, s}$ is
		\begin{equation}   \label{max_g_cdf}
			F_{\bar{g}_{m, s}}(x) = P(\bar{g}_{m, s} \leq x) = \prod_{n = 1}^{N} F_{|g_{n, m, s}|^2}(x) = (1 - e^{-x})^N.
		\end{equation}
	Furthermore, the probability density function (PDF) of $\bar{g}_{m, s}$ can be calculated as
		\begin{align} \label{max_g_pdf}
			f_{\bar{g}_{m, s}}(x) =& \frac{d F_{\bar{g}_{m, s}}(x)}{d x}  \nonumber \\
			=& \left\{
			\begin{array}{lc}
				N e^{-x} (1 - e^{-x})^{N - 1}, &\mbox{$x \geq 0$}, \\
				0, &\mbox{otherwise.}
			\end{array} \right.
		\end{align}
	
	Let $p_{\text{IT}}$ denote the proportion of subcarriers allocated to IT devices in the entire time horizon. According to the offline optimal solution, we have
		\begin{equation}
				p_{\text{IT}} = \frac{M S - dT^*}{M S}.
		\end{equation}
	To mimic the offline optimal subcarrier allocation, we can derive a threshold of $\{\bar{g}_{m, s} \}$, denoted by $q$, such that the expected proportion of subcarriers allocated to IT is $p_{\text{IT}}$. Specifically, a subcarrier is allocated to IT if the corresponding $\bar{g}_{m, s}$ exceeds the threshold $q$ and allocated to over-the-air model aggregation otherwise. To this end, we employ the \emph{quantile function} of $\bar{g}_{m, s}$ to compute $q$ as
		\begin{align} \label{opt_q}
			q = Q_{\bar{g}_{m, s}}(1 - p_{\text{IT}}) =& \inf \{x \in \mathbb{R}: 1 - p_{\text{IT}} \leq F_{\bar{g}_{m, s}}(x)\}  \nonumber \\
			=& -\ln \left(1 - \sqrt[N]{1 - p_{\text{IT}}} \right).
		\end{align}
	Then, the online subcarrier allocation solution to (P2) is given by
	\begin{equation}  \label{opt_sa1}
		o_{m, s} = \left\{
		\begin{array}{lc}
			1, &\mbox{$\bar{g}_{m, s} < q$}, \\
			0, &\mbox{otherwise,}
		\end{array} \right.
	\end{equation}
	and
	\begin{equation}  \label{opt_sa2}
		b_{n, m, s} =  \left\{
		\begin{array}{lc}
			1, &\mbox{\!\!\!\!\!  $\bar{g}_{m, s} \geq q$ and $n = \argmax\limits_{n' \in \mathcal{N}} |g_{n', m, s}|^2$}, \\
			0, &\mbox{ otherwise.}
		\end{array} \right.
	\end{equation}

	We summarize the proposed threshold-based online subcarrier allocation algorithm in Algorithm \ref{third_algorithm}. We initialize $o_{m, s} = 1$, $b_{n, m, s} = 0, \forall n$ for all subcarriers and introduce an auxiliary variable $I_{\textrm{total}}$ to denote the accumulated number of subcarriers allocated to IT. If the number of allocated subcarriers for FL achieves its maximum value $dT^*$, the remaining RBs will be allocated to IT; see Lines 7-8. By doing this, the proposed algorithm allocates exactly $dT^*$ subcarriers to over-the-air FL, guaranteeing that the solution is always feasible to Problem (P2). Since the complexity of Lines 6-12 is $O(N)$, the computational complexity of allocating each subcarrier is $O(N)$. Therefore, the overall complexity of Algorithm \ref{third_algorithm} is $O(M S N)$.
	
	\begin{algorithm}[t]
			\caption{Threshold-Based Online Subcarrier Allocation Algorithm}  \label{third_algorithm}
			\begin{algorithmic}[1]
					\STATE \textbf{Input:} $T^*$; $d$; $S$; $M$; $N$;
					\STATE \textbf{Initialization:} $o_{m, s} = 1, \forall m, s$; $b_{n, m, s} = 0, \forall n, m, s$; $p_{\text{IT}} = \frac{M S - dT^*}{M S}$; $q = -\ln (1 - \sqrt[N]{1 - p_{\text{IT}}} )$; $I_{\textrm{total}} = 0$;
					\FOR{$s = 1, \cdots, S$}
					\FOR{$m = 1, \cdots, M$}
					\IF{$I_{\textrm{total}} < M S \cdot p_{\text{IT}}$}
					\STATE Compute $\bar{g}_{m, s} = \max_{n \in \mathcal{N}} |g_{n, m, s}|^2$ and $n = \argmax_{n' \in \mathcal{N}} |g_{n', m, s}|^2$;
					\IF{$(s - 1) M + m - I_{\textrm{total}} > dT^*$}
					\STATE Update $o_{m, s} = 0$ and $b_{n, m, s} = 1$;
					\ELSE
					\STATE Update $o_{m, s}$ and $b_{n, m, s}$ by \eqref{opt_sa1} and \eqref{opt_sa2};
					\ENDIF
					\STATE Update $I_{\textrm{total}} = I_{\textrm{total}} + (1 - o_{m, s})$;
					\ENDIF				
					\ENDFOR
					\ENDFOR
					\STATE \textbf{Output:} $\{b_{n, m, s}, o_{m, s} \}$.
			\end{algorithmic}
	\end{algorithm}
	
	\subsection{Performance Analysis}
	In this subsection, we analyze the performance of the proposed threshold-based online algorithm, i.e., Algorithm \ref{third_algorithm}. Let $\hat{g}_{m, s}$ denote the effective IT channel gain on the $m$-th subcarrier of the $s$-th OFDM symbol with the proposed threshold-based subcarrier allocation as
	\begin{equation}  \label{effective_g}
		\hat{g}_{m, s} = \left\{
		\begin{array}{lc}
			\bar{g}_{m, s}, &\mbox{$\bar{g}_{m, s} \geq q$}, \\
			0, &\mbox{otherwise.}
		\end{array} \right.
	\end{equation}
	With \eqref{effective_g}, the expected objective value achieved by Algorithm \ref{third_algorithm} is
		\begin{align}  \label{proposed_rate}
			&\mathbb{E} [R_{\text{avg}}]  \nonumber \\
			=& \mathbb{E} \bigg[ \frac{1}{M S} \sum_{s = 1}^S \sum_{m = 1}^M \sum_{n = 1}^N b_{n, m, s} \log_2 \bigg( 1 + \frac{P_{2} |g_{n, m, s}|^2}{\phi \sigma^2} \bigg) \bigg]  \nonumber \\
			=& p_{\text{IT}} \mathbb{E} \left[ \log_2 \left( 1 + \theta \hat{g}_{m, s} \right) \right],
		\end{align}
	where $\theta \triangleq \frac{P_2}{\phi \sigma^2}$ is a known constant. The following proposition gives a closed-form expression of \eqref{proposed_rate}.
		\begin{proposition} \label{proposition1}
			The expected average IT sum-rate achieved by Algorithm \ref{third_algorithm} is given by
				\begin{align}  \label{proposition1_eq}
					\mathbb{E} [R_{\text{avg}}] =& N \sum_{i = 0}^{N - 1} \frac{\binom{N - 1}{i} (-1)^i }{(i + 1) \ln 2} \bigg[ \ln \left(1 + \theta q \right) e^{- (i + 1) q}  \nonumber \\
					&+ e^{\frac{i + 1}{\theta}} E_1 \left( \frac{i + 1}{\theta} + (i + 1) q \right) \bigg],
				\end{align}
			where $q$ is given by \eqref{opt_q}, and $E_1 (z) \triangleq \int_{z}^{\infty} \frac{e^{-t}}{t} dt$ is the first-order exponential integral.
		\end{proposition}
		\begin{IEEEproof}
			See Appendix \ref{proposition1_proof}.
		\end{IEEEproof}

	We compare the performance of the proposed algorithm with a baseline called random subcarrier allocation (RSCA). Specifically, RSCA randomly allocates each subcarrier to IT with probability $p_{\text{IT}}$ and to over-the-air model aggregation with probability $1 - p_{\text{IT}}$. This results in a suboptimal solution to Problem (P2). We denote the achievable expected objective value of RSCA by
		\begin{equation}   \label{RSCA_rate}
			\mathbb{E} \big[\hat{R}_{\text{avg}}\big] = p_{\text{IT}} \mathbb{E} \left[ \log_2 \left( 1 + \theta \bar{g}_{m, s} \right) \right],
		\end{equation}
	and the rate improvement of the proposed algorithm by $\varrho(q) \triangleq \mathbb{E} [R_{\text{avg}}] - \mathbb{E} \big[\hat{R}_{\text{avg}}\big]$. The following proposition gives a closed-form expression of $\varrho(q)$.
		\begin{proposition} \label{proposition2}
			The expected average IT sum-rate in \eqref{RSCA_rate} is given by
			\begin{align}  \label{proposition2_eq1}
				&\mathbb{E} \big[\hat{R}_{\text{avg}}\big]  \nonumber \\
				=& \left[1 - (1 - e^{-q})^N \right] N \sum_{i = 0}^{N - 1} \frac{\binom{N - 1}{i} (-1)^i }{(i + 1) \ln 2}  e^{\frac{(i + 1)}{\theta}} E_1 \left( \frac{i + 1}{\theta} \right).
			\end{align}
			The rate improvement $\varrho(q)$ is always non-negative with the following closed-form expression:
			\begin{align}  \label{proposition2_eq2}
					\varrho(q) =& N \sum_{i = 0}^{N - 1} \frac{ \binom{N - 1}{i} (-1)^i }{(i + 1) \ln 2} \bigg[ \ln(1 + \theta q) e^{- (i + 1)q }  \nonumber \\
					&+ e^{\frac{i + 1}{\theta}} E_1 \left(\frac{(i + 1) (1 + \theta q)}{\theta} \right)  \nonumber \\
					&+ \left( (1 - e^{-q})^N - 1 \right) e^{\frac{i + 1}{\theta}} E_1 \left( \frac{i + 1}{\theta} \right) \bigg]  \nonumber \\
					\geq& 0.
			\end{align}
		\end{proposition}
		\begin{IEEEproof}
			See Appendix \ref{proposition2_proof}.
		\end{IEEEproof}

	\section{Numerical Results}
	 In this section, we conduct experiments to verify our theoretical analysis and evaluate the performance of the proposed design.
	
	\subsection{Experiment Settings}
		We simulate a single-cell model edge network that comprises one AP, $K = 20$ FL devices, and $N = 5$ IT devices with $M = 512$ OFDM subcarriers. The frequency-selective channel has an exponential power delay profile with a delay spread of $6$ taps. Unless otherwise stated, we set $S = 2000$ with each symbol duration equals to $16$ $\mu s$, $P_1 = P_2 = P = 1$ W, $\phi = 6$ dB, and $\sigma^2 = 0.1$ W.

	\begin{figure}[t]	
		\centering
		\includegraphics[scale=0.75]{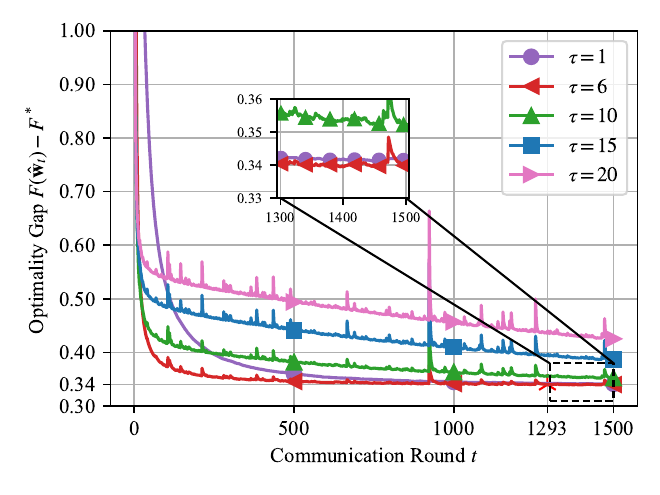}
		\caption{FL training optimality gap of different $\tau$.} \label{fig:tl_round}
	\end{figure}
 	\begin{figure}[t]	
 		\centering
 		\includegraphics[scale=0.75]{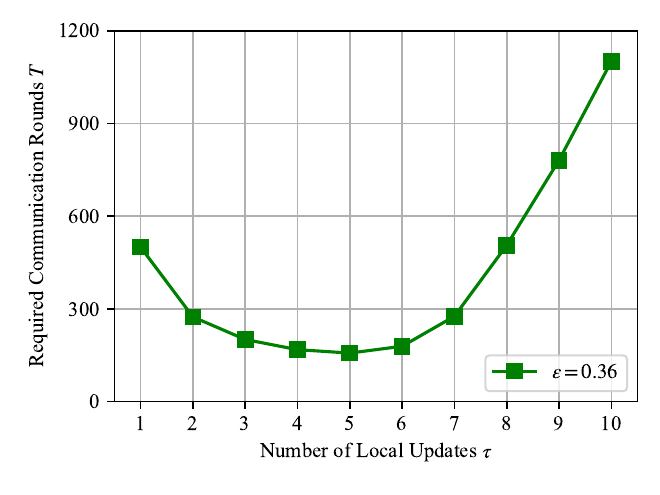}
 		\caption{Required $T$ to achieve an $\epsilon$-accurate solution.} \label{fig:req_T}
 	\end{figure}
	 	For over-the-air FL, we consider the regularized logistic regression task with the synthetic dataset specified in \cite[Section 5]{li2020federated}. Specifically, we generate the local data samples of each device $k$ $\{ \mathbf{X}_k, \mathbf{y}_k \}$ of size $D_k$ according to the model $y_{k, i} = \argmax (\text{softmax}(\mathbf{W}_k \mathbf{x}_{k, i} + \mathbf{b}_k))$, where $\text{softmax}(z_i) \triangleq \frac{e^{z_i}}{\sum_j e^{z_j}}$ is the softmax function, $\mathbf{X}_k \in \mathbb{R}^{D_k \times 60}$, $\mathbf{y}_k \in \mathbb{R}^{D_k  \times 1}$, $\mathbf{W}_k \in \mathbb{R}^{10 \times 60}$, $\mathbf{x}_{k, i} \in \mathbb{R}^{60 \times 1}$, and $\mathbf{b}_k \in \mathbb{R}^{10 \times 1}$. We draw each entry of $\mathbf{W}_k$ and $\mathbf{b}_k$ following the distribution of $\mathcal{N}(u_k, 1)$ with $u_k \sim \mathcal{N}(0, \alpha)$, and $(\mathbf{x}_{k, i})_j$ following the distribution of $\mathcal{N}(v_k, \frac{1}{j^{1.2}})$ with $v_k \sim \mathcal{N}(B_k, 1)$ and $B_k \sim \mathcal{N}(0, \beta)$. Here, $\alpha$ and $\beta$ affect the variances of the entries of $\mathbf{W}_k$, $\mathbf{b}_k$, and $\mathbf{x}_{k, i}$ and thus reflect the levels of model and data heterogeneities. In our simulations, we set $\alpha = \beta = 1$ by following \cite{li2019convergence}. The sizes of local datasets, i.e., $\{D_k\}$, are randomly drawn by following a power law distribution \cite{li2020federated}. The loss function is given by
		 	\begin{equation}  \label{simulation_loss}
		 		F(\mathbf{w}) = \frac{1}{D} \sum_{i = 1}^{D} \text{CrossEntropy} (f(\mathbf{w}; \mathbf{x}_i), y_i) + \frac{\varphi}{2} \|\mathbf{w}\|_2^2,
		 	\end{equation}
	 	where $(\mathbf{x}_i, y_i)$ is the $i$-th data sample, $f(\mathbf{w}; \mathbf{x}_i) \triangleq\text{softmax}$ $(\mathbf{W} \mathbf{x}_i + \mathbf{b})$ is the prediction model with the parameter $\mathbf{w} = (\mathbf{W}, \mathbf{b})$, and $\varphi = 0.5$ is the regularization parameter. To minimize \eqref{simulation_loss}, all the FL devices train a common regression model consisting of an input layer and a softmax output layer with the total number of model parameters $d$ equals to $610$. We set the mini-batch size $B = 32$, the local SGD clipping threshold $G = 1$, the learning rate $\lambda_t = 0.05 \cdot \frac{\gamma}{\gamma + t}$ and the weight of each global model iterate $\eta_t = (\gamma + t)^2$ with $\gamma =1000$. To verify the analysis, we numerically estimate the values of the learning hyper-parameters using centralized training and dataset statistics, which yields $\Gamma = 0.639, \mu = 0.5, L = 10.25$, and $\mathbb{E} [ \max_{k \in \mathcal{K}} \rho_k^2 / |h_k|^2] = 1.294$. All the simulation results in this section are averaged over $20$ Monte Carlo trials.

 	\subsection{Verification of Learning Analysis}
 	We first evaluate the optimality of the learning hyper-parameters $\tau^*$ and $T^*$ derived in Section III. In Fig. \ref{fig:tl_round}, we plot the optimality gap of over-the-air FL, i.e., $F(\hat{\mathbf{w}}_T) - F^*$, with different values of $\tau$ in $1,500$ communication rounds. Our analysis in \eqref{opt_tau} shows that the optimal $\tau^* = 6$, which corresponds to the best learning performance in Fig. \ref{fig:tl_round}.
 	We see that $\tau = 1$ leads to a slower convergence rate and a higher final training loss than $\tau = 6$. On the other hand, due to the data heterogeneity, we observe evident performance degradation when $\tau > 6$ as too intensive local updating hurts the global convergence. With $\tau = 6$, it requires $T = 1293$ rounds to achieve a $0.34$-accurate solution, which matches our analytical result $T^* = 1208$ in \eqref{opt_T}.

    \begin{table}[t]
		\caption{Average IT sum-rate $R_{\textrm{avg}}$ (Kbps) with $S=2000, N = 5$.}
		\centering
		\begin{tabular}{|c|c|c|c|c|c|}
			\hline
			Scheme & Offline & Proposed & RSCA & $\tau = 1$ & $\tau = 10$ \\
			\hline
			$R_{\textrm{avg}}$ (Kbps) & $66.40$ & $66.28$ & $52.38$ & $0$ & $53.04$ \\
			\hline
		\end{tabular}
		\label{table:tr_symbol}
	\end{table}
	
	\begin{figure}[t]	
		\centering
		\includegraphics[scale=0.75]{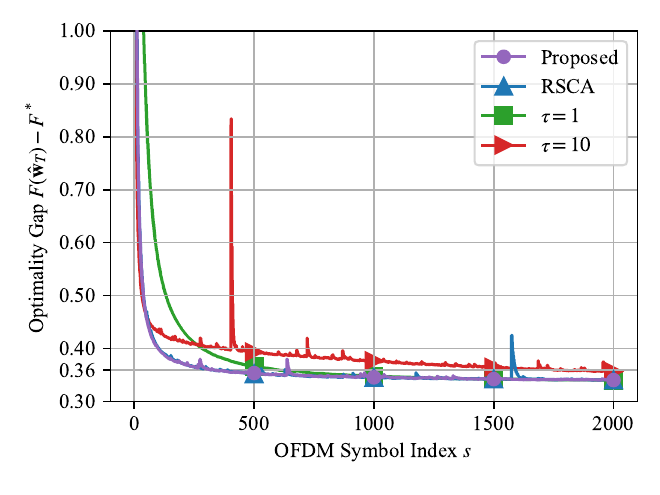}
		\caption{FL training optimality gap with $S=2000, N = 5$.} \label{fig:tl_symbol}
	\end{figure}
		
 	In Fig. \ref{fig:req_T}, we study the impact of the number of local updates on the required FL communication rounds $T$ to achieve a $0.36$-accurate solution, i.e., $F(\hat{\mathbf{w}}_T) - F^* \leq 0.36$. We see from Fig. \ref{fig:req_T} that the value of $T$ is sensitive to the choice of $\tau$ and their relationship is non-monotonic. This observation coincides with our analysis in Section III. Moreover, the optimal value of $\tau$ obtained by the simulation ($\tau^* = 5$) is close to our analysis ($\tau^* = 6$), which verifies the accuracy of our theoretical learning optimization results in Section III.

 	\begin{figure}[t]	
 		\centering
 		\includegraphics[scale=0.75]{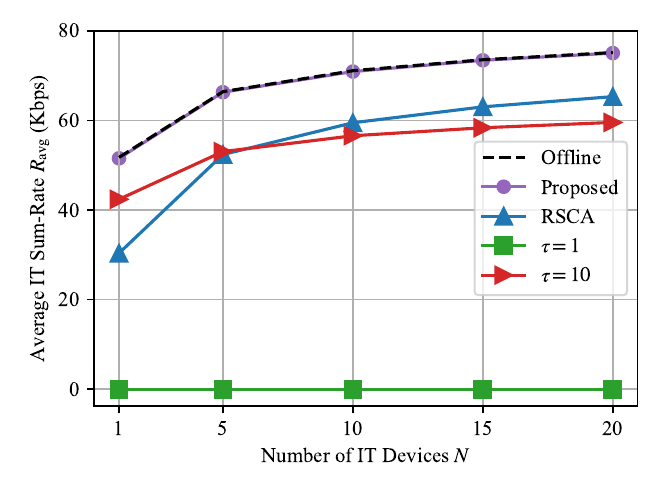}
 		\caption{Average IT sum-rate versus $N$ with $S=2000$.} \label{fig:tr_N}
 	\end{figure}	

 	\subsection{Simulations on CFLIT Performance}
 	We compare the proposed CFLIT design with four baseline schemes. First, we consider the offline optimal subcarrier allocation scheme described in Section IV-A, where we assume that the non-causal CSI is available beforehand. This ideal baseline characterizes the best IT performance. Second, we consider the RSCA scheme described in Section IV-B, which randomly allocates each RB to IT with probability $p_{\text{IT}}$ and to over-the-air FL with probability $1 - p_{\text{IT}}$. Then, we consider two FL schemes with fixed learning configuration $\tau = 1$ and $\tau = 10$, where the corresponding subcarrier allocation is optimized by using Algorithm \ref{third_algorithm}. For fair comparison, the numbers of local updates in the proposed scheme and RSCA are both optimized by \eqref{opt_tau} as $\tau^* = 6$. For over-the-air FL, the number of communication rounds is determined offline by \eqref{opt_T} with the learning target given by $F(\hat{\mathbf{w}}_T) - F^* \leq \epsilon = 0.36$.
 	
	In Table \ref{table:tr_symbol} and Fig. \ref{fig:tl_symbol}, we present the average IT sum-rate and the FL loss, respectively, with $S = 2000$ OFDM symbols. We observe that the proposed online algorithm achieves a comparable average IT sum-rate with the offline scheme. Moreover, the proposed scheme outperforms all the other baselines in terms of both the FL convergence and the IT data rate. Although the optimality gap of the baseline with $\tau = 1$ gradually approaches that of the proposed scheme, it fails to serve the IT devices as all the radio resources are used for over-the-air model aggregation. Due to the data heterogeneity, the baseline with $\tau = 10$ tends to converge to local optima and requires more FL communication rounds than the proposed scheme, which leads to a worse CFLIT system performance. Thanks to the optimization of $\tau$, the RSCA scheme achieves comparable learning performance with the proposed scheme, but it achieves a lower IT data rate than our proposed method. By joint communication and learning design, the proposed scheme efficiently utilizes the limited radio resources and thus provides better performance for the coexistence of over-the-air FL and IT.
	
	In Fig. \ref{fig:tr_N}, we study the impact of the number of IT devices $N$ on the average IT sum-rate with $S = 2000$ OFDM symbols. We see that the proposed scheme achieves a similar average IT sum-rate to the offline scheme and outperforms the other baselines. As $N$ increases, the performance gap between the RSCA scheme and the proposed scheme becomes smaller. This is because increasing the number of IT devices improves the effective received signal-to-noise ratio (SNR) for all the schemes. In contrast, the two baselines with fixed FL design, i.e., $\tau = 1$ and $\tau = 10$, suffer from a slow FL convergence rate. Consequently, they have much less remaining RBs for uplink IT compared with the proposed scheme.
 		
	\begin{figure}[t]	
		\centering
		\includegraphics[scale=0.75]{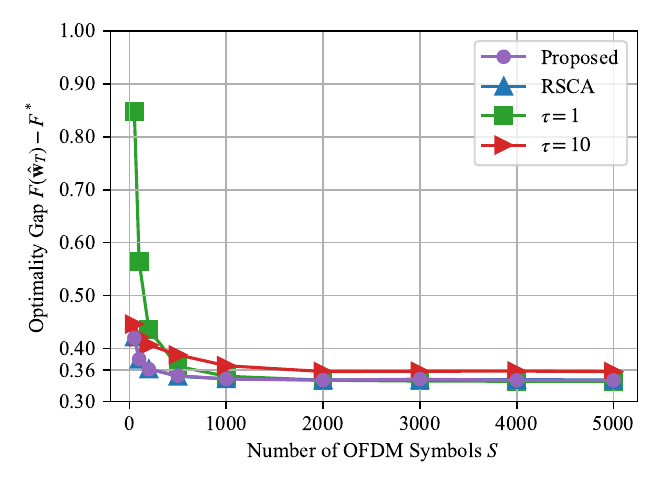}
		\caption{FL training optimality gap versus $S$ with $N = 5$.} \label{fig:tl_S}
	\end{figure}	
	\begin{figure}[t]	
		\centering
		\includegraphics[scale=0.75]{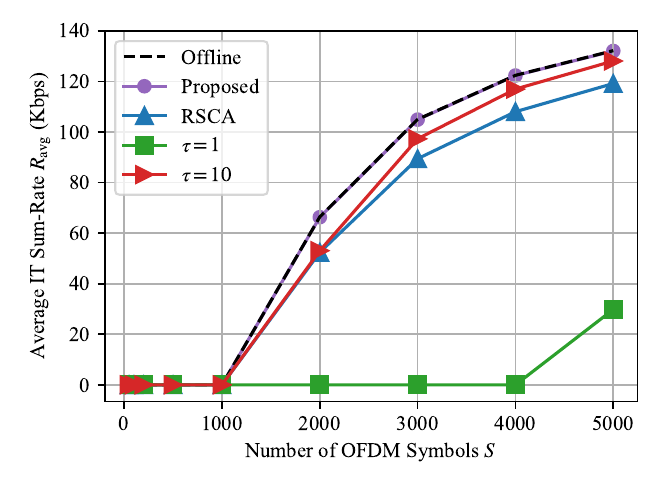}
		\caption{Average IT sum-rate versus $S$ with $N = 5$.} \label{fig:tr_S}
	\end{figure}	

	\begin{figure}[t]	
		\centering
		\includegraphics[scale=0.75]{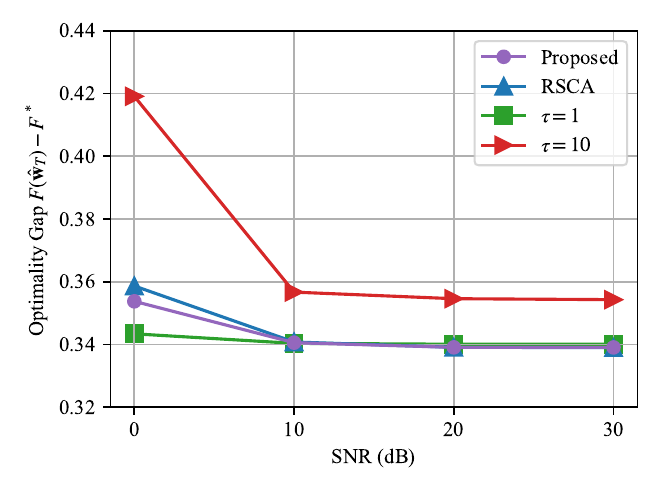}
		\caption{FL training optimality gap versus SNR with $S=2000, N = 5$.} \label{fig:tl_SNR}
	\end{figure}	
	
	\begin{figure}[t]	
		\centering
		\includegraphics[scale=0.75]{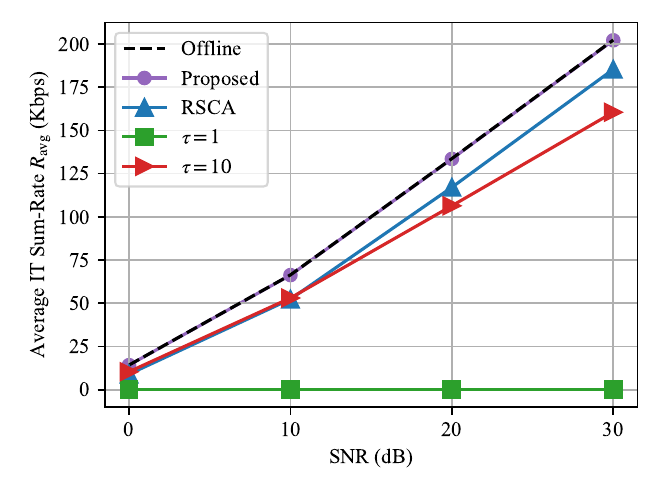}
		\caption{Average IT sum-rate versus SNR with $S=2000, N = 5$.} \label{fig:tr_SNR}
	\end{figure}

 	In Figs. \ref{fig:tl_S} and \ref{fig:tr_S}, we study the impact of the number of OFDM symbols $S$ on the FL performance and the average IT sum-rate, respectively. We see that the value of $S$ reflects the availability of radio resources and thus critically affects the performance of both the FL and the IT. Specifically, when $S$ is small, e.g., $S \leq 2000$, the radio resources should be allocated to over-the-air FL to fulfill the convergence requirement, which limits the data rate of IT devices. As $S$ increases, more RBs can be used for uplink data transmissions, leading to a higher IT data rate. As shown in Fig. \ref{fig:tr_S}, the proposed online algorithm achieves near-optimal IT performance. With efficient exploitation of the learning properties, our proposed design well balances the resource allocation in FL and IT and thus achieves a better performance compared with the other baselines. The above observation verifies the superiority of the proposed design in supporting the coexistence of FL and IT, especially when radio resources are limited.

	Next, we investigate the effect of the SNR, defined as the ratio of the transmit power $P$ to the noise power $\sigma^2$. Figs. \ref{fig:tl_SNR} and \ref{fig:tr_SNR} show that FL and IT exhibit distinct sensitivities to communication errors. Although a higher SNR leads to a larger IT data rate, its impact on FL performance is not consistently monotonic. For example, when the SNR decreases from 30 dB to 10 dB, the training performance is not affected because the model aggregation error is small enough to preserve convergence.  However, when the SNR falls below 10 dB, the substantial communication errors cause significant perturbations to the gradient directions, resulting in a larger training loss.

	\subsection{CFLIT Performance with Model Compression}
\begin{figure}[t]	
	\centering
	\includegraphics[scale=0.75]{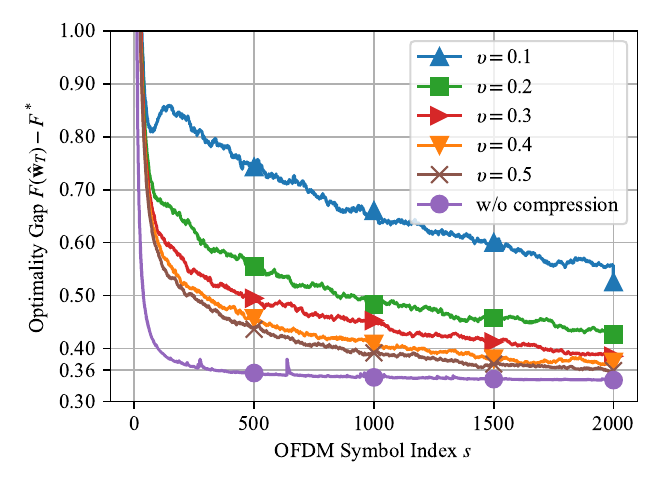}
	\caption{FL training optimality gap versus $\upsilon$ with $s=0.1d, S=2000, N = 5$.} \label{fig:tl_symbol_MC}
\end{figure}

	Finally, we show the feasibility of fusing the proposed method with model sparsification/compression. We simulate the proposed design with the existing model compression scheme in \cite{amiri2020machine}. Specifically, before uploading the local change vector, each FL device first sparsifies it to an $s$-sparse vector and then projects the sparsified vector to the $d^{\prime}$-dimensional subspace using an i.i.d. Gaussian projection matrix $\mathbf{A} \in \mathbb{R}^{d^{\prime} \times d}$, with $s \leq d^{\prime} < d$. The model compression ratio is given by $\upsilon \triangleq d^{\prime} / d \in (0, 1]$. Accordingly, the number of RBs for over-the-air model aggregation is changed from $dT^*$ to $\upsilon dT^*$. Therefore, $\upsilon$ reflects the communication-compression tradeoff: A smaller $\upsilon$ means more compressed and thus less accurate model aggregation but less radio resource consumption. Table \ref{table:tr_symbol_MC} and Fig. \ref{fig:tl_symbol_MC} show the average IT sum-rate and the FL training loss versus the value of $\upsilon$. The simulation parameters can be found in Fig. \ref{fig:tl_symbol}. Decreasing $\upsilon$ results in lower radio resource consumption during FL, which in turn increases the IT data rate. However, compressing the gradient vectors introduces additional errors to model aggregation and impairs the training performance. Particularly, when $\upsilon$ is 0.1, the compression error is significantly large, leading to a large training loss. The result demonstrates that the proposed design can be readily extended to the system with model sparsification/compression.
	
\begin{table*}[t]
	\centering
	\caption{Average IT sum-rate $R_{\textrm{avg}}$ (Kbps) versus $\upsilon$ with $s=0.1d, S=2000, N = 5$.}
	\begin{tabular}{|c|c|c|c|c|c|c|}
		\hline
		Compression ratio $\upsilon$ & $0.1$ & $0.2$ & $0.3$ & $0.1$ & $0.5$ & w/o compression \\
		\hline
		$R_{\textrm{avg}}$ (Kbps) & $157.31$ & $149.66$ & $141.17$ & $131.97$ & $122.29$ & $66.28$\\
		\hline
	\end{tabular}
	\label{table:tr_symbol_MC}
\end{table*}

\section{Conclusions}
	In this paper, we proposed a novel CFLIT framework to support the coexistence of over-the-air FL and uplink information transfer. To maximize the average IT data rate while ensuring the convergence of over-the-air FL, we studied the joint optimization of the OFDM subcarrier allocation, the transceiver design, and the FL hyper-parameters. We analyzed the convergence of over-the-air FL to minimize the bandwidth consumption needed for FL convergence guarantee. Based on the analysis, we optimized the transceiver design, the number of local updates, and the number of FL communication rounds. Moreover, we proposed a low-complexity online algorithm to optimize the CFLIT subcarrier allocation under the time-varying channels and derived a closed-form expression of its achievable IT data rate. Extensive simulations verify our analysis and demonstrate the performance improvement, in both the FL convergence and the average IT data rate, of the proposed design compared with the baselines.
	
	Finally, we conclude the paper with some interesting future directions. First, we assumed perfect synchronization of the local model updates transmitted from FL devices for over-the-air model aggregation, which is difficult to achieve in practice. The lack of synchronization will result in not only signal misalignment but also inter-symbol interference. Therefore, new designs on pre-equalization and de-noising are needed to address the synchronization issue in the coexisting system. Second, we showed numerically in Section V-D that combining our design with separate model compression can reduce the communication cost of FL. Joint optimization of the model aggregation frequency and model compression/quantization under a unified framework would further improve the communication efficiency in over-the-air FL. Moreover, device selection and scheduling are also key designing factors that significantly affect the FL convergence. Further investigation is necessary to fully understand their impacts on the performance of the coexisting system.

\appendices
	
\section{Proof of Lemma \ref{MMSE}} \label{MMSE_proof}
Substituting \eqref{aggregated_signal} into the expression of $e_t$, we have
\begin{align} \label{MMSE_eq1}
	e_{t} =& \mathbb{E}[\|\hat{\mathbf{r}}_t - \mathbf{r}_t \|_2^2]  \nonumber \\
	=& \sum_{i = 1}^d \mathbb{E} \Bigg[ \Bigg|\sum_{k = 1}^K \left(\frac{c_t[i] h_{k, \Phi_{t}[i]} p_{k, \Phi_{t}[i]}}{\nu_{k, t}} - \rho_k \right) \big(\Delta_{k, t}[i]  \nonumber \\
	& - \overline{\Delta}_{k, t} \big) \Bigg|^2 + |c_t[i] z_t[i]|^2 \Bigg]  \nonumber \\
	\overset{(a)}{\geq}& \sum_{i = 1}^d \mathbb{E}\left[ |c_t[i] z_t[i]|^2 \right] = \sum_{i = 1}^d |c_t[i]|^2 \sigma^2,
\end{align}
where the equality in $(a)$ holds if $\frac{c_t[i] h_{k, \Phi_{t}[i]} p_{k, \Phi_{t}[i]}}{\nu_{k, t}} - \rho_k = 0, \forall k \in \mathcal{K}$. From \cite[eqs. (49)-(54)]{lin2022relay}, we obtain the optimal $p_{k, \Phi_{t}[i]}$ and $c_t[i]$ as
\begin{equation}
		p_{k, \Phi_{t}[i]} = \frac{\rho_k \nu_{k, t}}{c_t[i] h_{k, \Phi_{t}[i]}}, \ \forall k \in \mathcal{K}, c_t[i] = \frac{1}{\sqrt{P_1}} \max_{k \in \mathcal{K}} \frac{\rho_k \nu_{k, t}}{|h_{k, \Phi_{t}[i]}|}. \label{MMSE_eq3}
\end{equation}
Plugging \eqref{MMSE_eq3} into \eqref{MMSE_eq1}, we obtain the minimum $e_t$ as $e_{t} = \frac{\sigma^2}{P_1} \sum_{i = 1}^d \max_{k \in \mathcal{K}} \frac{\rho_k^2 \nu_{k, t}^2}{|h_{k, \Phi_{t}[i]}|^2}$.

\section{Proof of Lemma \ref{MMSE2}}  \label{MMSE2_proof}	

By taking expectation with respect to the local mini-batch samples and the channel fading coefficients on both sides of \eqref{MMSE_eq} and replacing $h_{k, \Phi_{t}[i]}, \forall i,$ by $h_k$, we obtain
\begin{align} \label{expected_MSE}
		\mathbb{E} [ e_t ] =&\mathbb{E} \left[\frac{\sigma^2}{P_1} \sum_{i = 1}^d \max_{k \in \mathcal{K}} \frac{\rho_k^2 \nu_{k, t}^2}{|h_{k}|^2} \right] = \mathbb{E} \left[\frac{d \sigma^2}{P_1} \max_{k \in \mathcal{K}} \frac{\rho_k^2 \nu_{k, t}^2}{|h_{k}|^2} \right],
\end{align}
where the last equality is because $\{h_k\}$ are i.i.d. over the index $i$. For $\forall k \in \mathcal{K}$, we have
\begin{align}  \label{expected_MSE_SGD}
		\mathbb{E} \left[d \rho_k^2 \nu_{k, t}^2 \right] =& \mathbb{E} \left[\rho_k^2 \sum_{i = 1}^{d} \left(\Delta_{k, t}[i] - \frac{1}{d} \sum_{i' = 1}^{d} \Delta_{k, t}[i'] \right)^2 \right]  \nonumber \\
		=&\mathbb{E} \left[\rho_k^2 \left(\sum_{i = 1}^{d} \Delta_{k, t}^2[i] - \frac{1}{d} \left(\sum_{i' = 1}^{d} \Delta_{k, t}[i']\right)^2 \right) \right]  \nonumber \\
		\overset{(a)}{\leq}&\mathbb{E} \left[ \rho_k^2 \sum_{i = 1}^{d} \Delta_{k, t}^2[i] \right] \nonumber \\
		=& \mathbb{E} \left[ \rho_k^2 \left\|- \lambda_t \sum_{l = 1}^{\tau} \nabla F_k(\mathbf{w}_{k, t}^l, \xi_{k, t}^l) \right\|_2^2 \right] \nonumber \\
		\overset{(b)}{\leq}& \rho_k^2 \lambda_t^2 \tau \sum_{l = 1}^{\tau} \mathbb{E} \left[ \left\|\nabla F_k(\mathbf{w}_{k, t}^l, \xi_{k, t}^l) \right\|_2^2 \right]  \nonumber \\
		\overset{(c)}{\leq}& \rho_k^2 \lambda_t^2 \tau^2 G^2,
\end{align}
where $(a)$ is because $(\sum_{i' = 1}^{d} \Delta_{k, t}[i'])^2 / d \geq 0$, $(b)$ is from the convexity of $\|\cdot\|_2^2$, and $(c)$ is from \eqref{square_norm}. Plugging \eqref{expected_MSE_SGD} into \eqref{expected_MSE} gives \eqref{MMSE2_eq}, which completes the proof.

\section{Proof of Theorem \ref{theorem1}}  \label{theorem1_proof}	
	To prove Theorem \ref{theorem1}, we first introduce the following lemma.
	\begin{lemma} \label{lemma3}
		With Assumptions 1-3 and $\lambda_t \leq \frac{1}{2L \tau}$, we have
		\begin{align}  \label{lemma3_eq}
				&\mathbb{E} [\| \mathbf{w}_{t + 1} - \mathbf{w}^* \|_2^2]  \nonumber \\
				\leq& \left(1 - \frac{\mu \lambda_t \tau}{2} \right) \mathbb{E} \left[ \| \mathbf{w}_{t} - \mathbf{w}^* \|_2^2 \right] - \frac{1}{2} \lambda_t \tau \mathbb{E} \left[ F(\mathbf{w}_{t}) - F^* \right]  \nonumber \\
				&+ \lambda_t^2 \tau^2 G^2 + 2 \sum_{k = 1}^K \rho_k \sum_{l = 2}^{\tau} \mathbb{E} \left[ \|\mathbf{w}_{k, t}^{l} - \mathbf{w}_{t}\|_2^2 \right] + 4 \lambda_t^2 \tau L \Gamma  \nonumber \\
				&+ \mathbb{E} \left[ \| \hat{\mathbf{r}}_t - \mathbf{r}_t \|_2^2 \right].
		\end{align}
	\end{lemma}
	\begin{IEEEproof}
		See Appendix \ref{lemma3_proof}.
	\end{IEEEproof}

	Since $\mathbf{w}_{k, t}^{l} - \mathbf{w}_{t} = - \lambda_t \sum_{j = 1}^{l - 1} \nabla F_k(\mathbf{w}_{k, t}^{j}, \xi_{k, t}^j)$, we have
	\begin{align}  \label{theorem1_eq2}
			&\sum_{k = 1}^K \rho_k \sum_{l = 2}^{\tau} \mathbb{E} \left[ \|\mathbf{w}_{k, t}^{l} - \mathbf{w}_{t}\|_2^2 \right]  \nonumber\\
			=& \lambda_t^2 \sum_{k = 1}^K \rho_k \sum_{l = 2}^{\tau} \mathbb{E} \left[ \left\| \sum_{j = 1}^{l - 1} \nabla F_k(\mathbf{w}_{k, t}^{j}, \xi_{k, t}^j) \right\|_2^2 \right]  \nonumber \\
			\overset{(a)}{\leq}& \lambda_t^2 \sum_{k = 1}^K \rho_k \sum_{l = 2}^{\tau} (l - 1) \sum_{j = 1}^{l - 1}  \mathbb{E} \left[ \| \nabla F_k(\mathbf{w}_{k, t}^{j}, \xi_{k, t}^j) \|_2^2 \right]  \nonumber \\
			\overset{(b)}{\leq}& \lambda_t^2 \sum_{k = 1}^K \rho_k \sum_{l = 2}^{\tau} (l - 1)^2 G^2 = \lambda_t^2 \frac{\tau (\tau - 1) (2\tau - 1)}{6} G^2,
	\end{align}
	where $(a)$ is from the convexity of $\|\cdot\|_2^2$, and $(b)$ is from \eqref{square_norm}.

	Moreover, from Lemma \ref{MMSE2}, we have
	\begin{equation}  \label{theorem1_eq3}
			\mathbb{E} [ \| \hat{\mathbf{r}}_t - \mathbf{r}_t \|_2^2 ] \leq \lambda_t^2 \frac{\tau^2 G^2 \sigma^2}{P_1} \mathbb{E}\left[ \max_{k \in \mathcal{K}} \frac{\rho_k^2}{|h_{k}|^2} \right].
	\end{equation}
	Substituting \eqref{theorem1_eq2} and \eqref{theorem1_eq3} into \eqref{lemma3_eq}, we have
	\begin{align}  \label{theorem1_eq4}
			&\mathbb{E} [\| \mathbf{w}_{t + 1} - \mathbf{w}^* \|_2^2]  \nonumber \\
			\leq&\left(1 - \frac{\mu \lambda_t \tau}{2} \right) \mathbb{E} [\| \mathbf{w}_{t} - \mathbf{w}^* \|_2^2] - \lambda_t \frac{1}{2} \tau \mathbb{E} \left[F(\mathbf{w}_{t}) - F^*\right]  \nonumber \\
			&+ \lambda_t^2 \left( \frac{2\tau^3 + \tau}{3} G^2 + 4 \tau L \Gamma + \frac{\tau^2 G^2 \sigma^2}{P_1} \mathbb{E} \left[ \max_{k \in \mathcal{K}} \frac{\rho_k^2}{|h_k|^2} \right] \right).
	\end{align}

	Next, we introduce the following lemma to bound the expected value of the training loss.

	\begin{lemma} \label{lemma2}
	Suppose that two sequences $\{a_t: a_t \geq 0\}$ and $\{e_t: e_t \geq 0\}$ satisfy
	\begin{equation}  \label{lemma2_eq1}
		a_{t + 1} \leq \left(1 - \frac{\mu \lambda_t \tau}{2} \right) a_t - \lambda_t e_t A + \lambda_t^2 C,
	\end{equation}
	for $\lambda_t = \frac{8}{\mu \tau (\gamma + t)}$ and constants $A > 0$, $C \geq 0$, $\mu > 0$, $\gamma \geq 1$. Then, we have
	\begin{equation}  \label{lemma2_eq2}
		\frac{A}{S_T} \sum_{t = 0}^{T - 1} \eta_t e_t \leq \frac{\mu \gamma^3 \tau}{8 S_T} a_0 + \frac{4T(T + 2 \gamma)}{\mu \tau S_T} C,
	\end{equation}
	for $\eta_t = (\gamma + t)^2$ and $S_T \triangleq \sum_{t = 0}^{T - 1} \eta_t$.
	\end{lemma}
	\begin{IEEEproof}
		The result in \eqref{lemma2_eq2} follows from \cite[Lemma 3.4]{stich2018local}.
	\end{IEEEproof}

    By the convexity of $F(\cdot)$, we have
		\begin{align} \label{lemma2_eq3}
			\mathbb{E} [F(\hat{\mathbf{w}}_T) - F^*] =& \mathbb{E} \left[F \left(\frac{1}{S_T} \sum_{t = 0}^{T - 1} \eta_t \mathbf{w}_t \right) - F^* \right]  \nonumber \\
			\leq& \frac{1}{S_T} \sum_{t = 0}^{T - 1} \eta_t \mathbb{E} [F(\mathbf{w}_t) - F^*].
		\end{align}
	The definitions of $L$ and $\mu$ yield $\frac{L}{\mu} \geq 1$. Let $\gamma \geq \frac{16 L}{\mu} \geq 1$ be a given constant. We note that $\lambda_t = \frac{8}{\mu \tau (\gamma + t)}$ satisfies the condition of Lemma \ref{lemma3}, i.e., $\lambda_t \leq \frac{1}{2 L \tau}$. Therefore, substituting $a_t = \mathbb{E} [\| \mathbf{w}_{t} - \mathbf{w}^* \|_2^2]$, $e_t = \mathbb{E} \left[F(\mathbf{w}_{t}) - F^*\right]$, $A = \frac{1}{2} \tau$, and $C = \frac{2\tau^3 + \tau}{3} G^2 + 4 \tau L \Gamma + \frac{\tau^2 G^2 \sigma^2}{P_1} \mathbb{E} \big[ \max_{k \in \mathcal{K}} \frac{\rho_k^2}{|h_k|^2} \big]$ into \eqref{theorem1_eq4}, we obtain from Lemma \ref{lemma2} and \eqref{lemma2_eq3} that
		\begin{align} \label{result_lemma2}
			\frac{1}{2} \tau \mathbb{E} [F(\hat{\mathbf{w}}_T) - F^*] \leq& \frac{4T(T + 2 \gamma)}{\mu \tau S_T} \bigg( \frac{2\tau^3 + \tau}{3} G^2 + 4 \tau L \Gamma  \nonumber \\
			&+ \frac{\tau^2 G^2 \sigma^2}{P_1} \mathbb{E} \left[ \max_{k \in \mathcal{K}} \frac{\rho_k^2}{|h_k|^2} \right] \bigg)  \nonumber \\
			&+ \frac{\mu \gamma^3 \tau}{8 S_T} \mathbb{E} [\| \mathbf{w}_{0} - \mathbf{w}^* \|_2^2].
		\end{align}
	Dividing both sides of \eqref{result_lemma2} by $\frac{1}{2} \tau$, we have \eqref{convergence}.

\section{Proof of Proposition \ref{proposition1}} \label{proposition1_proof}
Since $\lim_{x \rightarrow \infty}F_{\hat{g}_{m, s}}(x) = 1$, we can compute the PDF of $\hat{g}_{m, s}$ as
	\begin{align}
		f_{\hat{g}_{m, s}}(x) =& \frac{f_{\bar{g}_{m, s}}(x) \cdot \bm{1}_{x \geq q}}{\int_{q}^{\infty} f_{\bar{g}_{m, s}}(x) dx} = \frac{f_{\bar{g}_{m, s}}(x) \cdot \bm{1}_{x \geq q}}{1 - F_{\bar{g}_{m, s}}(q)}  \nonumber \\
		=& \begin{cases}
			\frac{N e^{-x} (1 - e^{-x})^{N - 1}}{p_{\text{IT}}}, &\mbox{$x \geq q$}. \\
			0, &\mbox{otherwise.}
		\end{cases}
	\end{align}
Therefore, we have
	\begin{align}  \label{avg_rate}
		\mathbb{E} \left[R_{\text{avg}}\right] \!=& p_{\text{IT}} \mathbb{E} \left[ \log_2 \left( 1 + \theta \hat{g}_{m, s} \right) \right]  \nonumber \\
		=& p_{\text{IT}} \int_{-\infty}^{\infty} \log_2 \left( 1 + \theta x \right) f_{\hat{g}_{m, s}} (x) d x  \nonumber \\
		\overset{(a)}{=}& \int_{q}^{\infty} \log_2 \left( 1 + \theta x \right) N e^{-x } \sum_{i = 0}^{N - 1} \binom{N - 1}{i} (-e^{-x})^i d x  \nonumber \\
		=& N \sum_{i = 0}^{N - 1} \frac{\binom{N - 1}{i} (-1)^i}{\ln 2}  \int_{q}^{\infty} \ln \left( 1 + \theta x \right) e^{-(i + 1) x} d x,
	\end{align}
where $(a)$ is from the binomial expansion. Moreover, we have the following integral formula \cite[eq. (3.352.2), p. 340]{gradshteyn2007table}
	\begin{align}  \label{integral}
		&\int_{a}^{\infty} \ln(1 + \vartheta) e^{- \omega \vartheta} d\vartheta \nonumber \\
		=& \ln(1 + a) \frac{e^{- \omega a}}{\omega} + \frac{1}{\omega} \int_{a}^{\infty}  \frac{e^{- \omega \vartheta}}{1 + \vartheta} d\vartheta  \nonumber \\
		=& \frac{1}{\omega} \left[\ln(1 + a) e^{- \omega a} + e^{\omega} E_1 (\omega a + \omega) \right].
	\end{align}
Substituting \eqref{integral} into \eqref{avg_rate} with $\vartheta = \theta x$ and $\omega = \frac{i + 1}{\theta}$, we obtain the result in \eqref{proposition1_eq}.

\section{Proof of Proposition \ref{proposition2}} \label{proposition2_proof}
	From \eqref{max_g_pdf}, we have
		\begin{align}  \label{hat_avg_rate}
			&\mathbb{E} \big[\hat{R}_{\text{avg}}\big]  \nonumber \\
			=& p_{\text{IT}} \mathbb{E} \left[ \log_2 ( 1 + \theta \bar{g}_{m, s} ) \right]  \nonumber \\
			=& p_{\text{IT}} \int_{-\infty}^{\infty} \log_2 ( 1 + \theta x ) f_{\bar{g}_{m, s}} (x) d x  \nonumber \\
			\overset{(a)}{=}& p_{\text{IT}} \int_{0}^{\infty} \log_2 \left(1 \! + \! \theta x \right) N e^{-x } \sum_{i = 0}^{N - 1} \binom{N - 1}{i} (-e^{-x})^i d x	 \nonumber \\
			=& p_{\text{IT}} N \sum_{i = 0}^{N - 1} \frac{\binom{N - 1}{i} (-1)^i }{(i + 1) \ln 2} e^{\frac{i + 1}{\theta}} E_1 \left(\! \frac{i + 1}{\theta} \!\right),
		\end{align}
	where $(a)$ is from the binomial expansion, and the last equality follows from \eqref{avg_rate} and \eqref{integral} with $a = 0$, $\vartheta = \theta x$ and $\omega = \frac{i + 1}{\theta}$. Note that the condition $q = -\ln \left(1 - \sqrt[N]{1 - p_{\text{IT}}} \right)$ is equivalent to
		\begin{equation}  \label{p_it}
			p_{\text{IT}} = 1 - (1 - e^{-q})^N.
		\end{equation}
	Plugging \eqref{p_it} into \eqref{hat_avg_rate}, we have
		\begin{align}  \label{hat_avg_rate2}
			\mathbb{E} \big[\hat{R}_{\text{avg}}\big] =& N \sum_{i = 0}^{N - 1} \frac{\binom{N - 1}{i} (-1)^i }{(i + 1) \ln 2} \bigg[ e^{\frac{i + 1}{\theta}} E_1 \left( \frac{i + 1}{\theta} \right)  \nonumber \\
			&- (1 - e^{-q})^N e^{\frac{i + 1}{\theta}} E_1 \left( \frac{i + 1}{\theta} \right) \bigg].
		\end{align}
	Note that $\varrho(q) \triangleq \mathbb{E} [R_{\text{avg}}] - \mathbb{E} \big[\hat{R}_{\text{avg}}\big]$. By \eqref{proposition1_eq} and \eqref{hat_avg_rate2}, we have
		\begin{align}  \label{func_q}
			\varrho(q) =& N \sum_{i = 0}^{N - 1} \frac{ \binom{N - 1}{i} (-1)^i }{(i + 1) \ln 2} \bigg[ \ln(1 + \theta q) e^{- (i + 1)q }  \nonumber \\
			&+ e^{\frac{i + 1}{\theta}} E_1 \left( \frac{i + 1}{\theta} + (i + 1) q \right) - e^{\frac{i + 1}{\theta}} E_1 \left( \frac{i + 1}{\theta} \right)  \nonumber \\
			& + (1 - e^{-q})^N e^{\frac{i + 1}{\theta}} E_1 \left( \frac{i + 1}{\theta} \right)  \bigg].
		\end{align}
	Taking the derivative of $\varrho(q)$, we have
		\begin{align}  \label{derivative_q}
			&\varrho'(q)  \nonumber\\
			=& N \sum_{i = 0}^{N - 1} \frac{ \binom{N - 1}{i} (-1)^i }{(i + 1) \ln 2} N e^{-q} (1 - e^{-q})^{N - 1} e^{\frac{i + 1}{\theta}} E_1 \left( \frac{i + 1}{\theta} \right)  \nonumber \\
			&- \frac{N \ln(1 + \theta q) e^{- q }}{\ln 2} \sum_{i = 0}^{N - 1} \binom{N - 1}{i} (-e^{- q })^i  \nonumber \\
			\overset{(b)}{=}& \frac{N e^{-q} (1 - e^{-q})^{N - 1}}{\ln 2} \Bigg[ N \sum_{i = 0}^{N - 1} \frac{ \binom{N - 1}{i} (-1)^i }{i + 1} e^{\frac{i + 1}{\theta}} E_1 \left(\! \frac{i + 1}{\theta} \! \right)  \nonumber \\
			&- \ln(1 + \theta q) \Bigg]  \nonumber \\
			\overset{(c)}{=}& \frac{N e^{-q} (1 - e^{-q})^{N - 1}}{\ln 2} \left[ \mathbb{E} \left[ \ln \left( 1 + \theta \bar{g}_{m, s} \right) \right] - \ln (1 + \theta q) \right],
		\end{align}
	where we have applied the property $E_1' (z) = - \frac{e^{-z}}{z}$, $(b)$ is from the binomial expansion, and $(c)$ is from \eqref{hat_avg_rate}.
	
	Define $R \triangleq \mathbb{E} \left[ \ln \left( 1 + \theta \bar{g}_{m, s} \right) \right]$. By setting $\varrho'(q^*) = 0$, we have
		\begin{equation}
			R - \ln (1 + \theta q^*) = 0  \quad \Rightarrow \quad q^* = \frac{e^R - 1}{\theta}.
		\end{equation}
	By \eqref{max_g_pdf}, we have $\mathbb{E}[\bar{g}_{m, s}] > 0$. Since $\theta = \frac{P_2}{\phi \sigma^2} > 0$, $R > 0$ and thus $q^* > 0$. It follows from \eqref{derivative_q} that $\varrho'(q) > 0$ when $0 \leq q < q^*$ and $\varrho'(q) < 0$ when $q > q^*$. Moreover, it is observed from \eqref{func_q} that $\varrho(0) = \lim_{q \rightarrow \infty} \varrho(q) = 0$. Therefore, we conclude that $\varrho(q) \geq 0$ for all $q$.

\section{Proof of Lemma \ref{lemma3}}  \label{lemma3_proof}
Let $\mathbf{v}_{t + 1} \in \mathbb{R}^{d \times 1}$ be the \emph{virtual} global model vector obtained from model aggregation without communication error in the $t$-th communication round. That is, $\mathbf{v}_{t + 1}$ is given by
	\begin{equation} \label{lemma3_proof_eq1}
		\mathbf{v}_{t + 1} = \mathbf{w}_{t} + \mathbf{r}_t.
	\end{equation}
Then, we have
	\begin{align} \label{lemma3_proof_eq2}
		&\mathbb{E} \left[ \| \mathbf{w}_{t + 1} - \mathbf{w}^* \|_2^2 \right]  \nonumber \\
		=& \mathbb{E} \left[ \| \mathbf{w}_{t + 1} - \mathbf{v}_{t + 1} \|_2^2 \right] + \mathbb{E} \left[ \| \mathbf{v}_{t + 1} - \mathbf{w}^* \|_2^2 \right]  \nonumber \\
		&+ 2 \mathbb{E} \left[ \langle \mathbf{w}_{t + 1} - \mathbf{v}_{t + 1}, \mathbf{v}_{t + 1} - \mathbf{w}^* \rangle \right]  \nonumber \\
		=& \mathbb{E} \left[ \| \hat{\mathbf{r}}_t - \mathbf{r}_t \|_2^2 \right] + \mathbb{E} \left[ \| \mathbf{v}_{t + 1} - \mathbf{w}^* \|_2^2 \right]  \nonumber \\
		&+ 2 \mathbb{E} \left[ \langle \hat{\mathbf{r}}_t - \mathbf{r}_t, \mathbf{v}_{t + 1} - \mathbf{w}^* \rangle \right],
	\end{align}
where the last equality is because $\mathbf{w}_{t + 1} - \mathbf{v}_{t + 1} = \mathbf{w}_{t} + \hat{\mathbf{r}}_t - \mathbf{w}_{t} - \mathbf{r}_t = \hat{\mathbf{r}}_t - \mathbf{r}_t$.

Note that
	\begin{align}
			\mathbb{E} \left[ \hat{r}_t[i] - r_t[i] \right] =& \mathbb{E} \Bigg[ \sum_{k = 1}^K \left(\frac{c_t[i] h_{k, \Phi_{t}[i]} p_{k, \Phi_{t}[i]}}{\nu_{k, t}} - \rho_k \right) \! \big(\Delta_{k, t}[i]  \nonumber \\
			&- \overline{\Delta}_{k, t} \big) + c_t[i] z_t[i] \Bigg]  \nonumber \\
			=& 0,
	\end{align}
where the last equality is because $\frac{c_t[i] h_{k, \Phi_{t}[i]} p_{k, \Phi_{t}[i]}}{\nu_{k, t}} - \rho_k = 0$, which is achieved by substituting the optimal $\{p_{k, \Phi_{t}[i]}, c_t[i]\}$ given in \eqref{opt_p_c}, and $\mathbb{E} \left[ z_t[i] \right] = 0$. Consequently,  we have $\mathbb{E} \left[ \hat{\mathbf{r}}_t - \mathbf{r}_t \right] = 0$ and thus $\mathbb{E} \left[ \langle \hat{\mathbf{r}}_t - \mathbf{r}_t, \mathbf{v}_{t + 1} - \mathbf{w}^* \rangle \right] = 0$.

Since $\mathbf{v}_{t + 1} = \mathbf{w}_{t} + \mathbf{r}_t$, we have
	\begin{align} \label{lemma3_proof_eq3}
			\mathbb{E} \left[ \| \mathbf{v}_{t + 1} - \mathbf{w}^* \|_2^2 \right] =& \mathbb{E} \left[ \| \mathbf{w}_{t} - \mathbf{w}^* \|_2^2 \right] + \mathbb{E} \left[ \| \mathbf{r}_t \|_2^2 \right]  \nonumber \\
			&+ 2 \mathbb{E} \left[ \langle \mathbf{w}_{t} - \mathbf{w}^*, \mathbf{r}_t \rangle \right].
	\end{align}
By the convexity of $\| \cdot \|_2^2$ and Assumption 3, we have
	\begin{align} \label{lemma7_eq2}
			\mathbb{E} \left[ \| \mathbf{r}_t \|_2^2 \right] =& \mathbb{E} \left[ \left\| - \lambda_t \sum_{k = 1}^{K} \rho_k \sum_{l = 1}^{\tau} \nabla F_k(\mathbf{w}_{k, t}^{l}, \xi_{k, t}^l) \right\|_2^2 \right]	 \nonumber \\
			\leq& \lambda_t^2 \sum_{k = 1}^K \rho_k \mathbb{E} \left[ \left\| \sum_{l = 1}^{\tau} \nabla F_k(\mathbf{w}_{k, t}^{l}, \xi_{k, t}^l) \right\|_2^2 \right]  \nonumber \\
			\leq& \lambda_t^2 \tau^2 G^2.
	\end{align}
Moreover, we have
	\begin{align} \label{lemma7_cross_term}
			&\mathbb{E} \left[ \langle \mathbf{w}_{t} - \mathbf{w}^*, \mathbf{r}_t \rangle \right]  \nonumber \\
			=& - \lambda_t \sum_{k = 1}^K \rho_k \sum_{l = 1}^{\tau}  \mathbb{E} \left[ \langle \mathbf{w}_{t} - \mathbf{w}^*, \nabla F_k(\mathbf{w}_{k, t}^{l}, \xi_{k, t}^l) \rangle \right]  \nonumber \\
			=& - \lambda_t \sum_{k = 1}^K \rho_k \sum_{l = 1}^{\tau}  \mathbb{E} \left[\langle \mathbf{w}_{t} - \mathbf{w}_{k, t}^{l} + \mathbf{w}_{k, t}^{l} - \mathbf{w}^*, \nabla F_k(\mathbf{w}_{k, t}^{l}) \rangle \right]  \nonumber \\
			=& - \lambda_t \sum_{k = 1}^K \rho_k \sum_{l = 2}^{\tau} \mathbb{E} \left[ \langle \mathbf{w}_{t} - \mathbf{w}_{k, t}^{l}, \nabla F_k(\mathbf{w}_{k, t}^{l}) \rangle \right]  \nonumber \\
			&- \lambda_t \sum_{k = 1}^K \rho_k \sum_{l = 1}^{\tau} \mathbb{E} \left[ \langle \mathbf{w}_{k, t}^{l} - \mathbf{w}^*, \nabla F_k(\mathbf{w}_{k, t}^{l}) \rangle \right],
	\end{align}
where the second equality is because $\mathbb{E} [\nabla F_k(\mathbf{w}_{k, t}^l, \xi_{k, t}^l)] = \nabla F_k(\mathbf{w}_{k, t}^l)$ due to the uniform randomness of local stochastic gradients, and the last equality is from the fact that $\mathbf{w}_{k, t}^{1} = \mathbf{w}_{t}$.
	
By the Cauchy-Schwarz inequality and the inequality of arithmetic and geometric means (AM-GM inequality), we have
	\begin{align} \label{AM-GM}
			-2 \langle \mathbf{w}_{t} - \mathbf{w}_{k, t}^{l}, \nabla F_k(\mathbf{w}_{k, t}^{l}) \rangle \leq& \frac{1}{\lambda_t} \|\mathbf{w}_{t} - \mathbf{w}_{k, t}^{l}\|_2^2  \nonumber \\
			&+ \lambda_t \|\nabla F_k(\mathbf{w}_{k, t}^{l}) \|_2^2.
	\end{align}
By the $\mu$-strong convexity of $F_k(\cdot)$, we have
	\begin{align} \label{lemma7_cvx}
			- \langle \mathbf{w}_{k, t}^{l} - \mathbf{w}^*, \nabla F_k(\mathbf{w}_{k, t}^{l}) \rangle \leq& - \left( F_k(\mathbf{w}_{k, t}^{l}) - F_k(\mathbf{w}^*) \right)  \nonumber \\
			&- \frac{\mu}{2} \|\mathbf{w}_{k, t}^{l} - \mathbf{w}^*\|_2^2.
	\end{align}
By the $L$-smoothness of $F_k(\cdot)$, we have
	\begin{equation} \label{smoothness}
			\| \nabla F_k(\mathbf{w}_{k, t}^{l}) \|_2^2 \leq 2L (F_k(\mathbf{w}_{k, t}^{l}) - F_k^*).
	\end{equation}
Since $\|\bm{a} + \bm{b} \|_2^2 \leq 2 \| \bm{a} \|_2^2 + 2 \| \bm{b} \|_2^2$, we further have
	\begin{equation} \label{lemma7_tri}
			- \|\mathbf{w}_{k, t}^{l} - \mathbf{w}^*\|_2^2 \leq \| \mathbf{w}_{t} - \mathbf{w}_{k, t}^{l} \|_2^2 - \frac{1}{2} \|\mathbf{w}_{t} - \mathbf{w}^*\|_2^2.
	\end{equation}
Combining \eqref{lemma3_proof_eq3}-\eqref{lemma7_tri},  we obtain
	\begin{align} \label{lemma7_decomp2}
			&\mathbb{E} \left[ \| \mathbf{v}_{t + 1} - \mathbf{w}^* \|_2^2 \right]  \nonumber \\
			\leq& \mathbb{E} \left[ \| \mathbf{w}_{t} - \mathbf{w}^* \|_2^2 \right] + \lambda_t^2 \tau^2 G^2  \nonumber \\
			&\hspace{-0.5em} + \! \lambda_t \sum_{k = 1}^K \rho_k \sum_{l = 2}^{\tau} \mathbb{E} \left[ \frac{1}{\lambda_t} \|\mathbf{w}_{t} - \mathbf{w}_{k, t}^{l}\|_2^2 + \lambda_t \|\nabla F_k(\mathbf{w}_{k, t}^{l})\|_2^2 \right]  \nonumber \\
			&\hspace{-0.5em} - \! 2 \lambda_t \sum_{k = 1}^K \! \rho_k \! \sum_{l = 1}^{\tau} \mathbb{E} \left[ F_k(\mathbf{w}_{k, t}^{l}) - F_k(\mathbf{w}^*) + \frac{\mu}{2} \|\mathbf{w}_{k, t}^{l} - \mathbf{w}^*\|_2^2 \right]  \nonumber \\
			\overset{(a)}{\leq}& \mathbb{E} \left[ \| \mathbf{w}_{t} - \mathbf{w}^* \|_2^2 \right] + \lambda_t^2 \tau^2 G^2 \! + \sum_{k = 1}^K \rho_k \sum_{l = 2}^{\tau} \mathbb{E} \left[ \| \mathbf{w}_{t} - \mathbf{w}_{k, t}^{l} \|_2^2 \right] \nonumber \\
			&\hspace{-0.5em} + \lambda_t^2 \sum_{k = 1}^K \rho_k \sum_{l = 2}^{\tau} \mathbb{E} \left[ 2 L (F_k(\mathbf{w}_{k, t}^{l}) - F_k^*) \right]  \nonumber \\
			&\hspace{-0.5em} - 2 \lambda_t \sum_{k = 1}^K \rho_k \sum_{l = 1}^{\tau} \mathbb{E} \left[ F_k(\mathbf{w}_{k, t}^{l}) - F_k(\mathbf{w}^*) \right]  \nonumber \\
			&\hspace{-0.5em} + \mu \lambda_t \sum_{k = 1}^K \rho_k \sum_{l = 1}^{\tau}  \left( \! \mathbb{E} \left[ \| \mathbf{w}_{t} - \mathbf{w}_{k, t}^{l} \|_2^2 \right] - \frac{1}{2} \mathbb{E} \left[ \|\mathbf{w}_{t} - \mathbf{w}^*\|_2^2 \right] \! \right)  \nonumber \\
			\overset{(b)}{\leq}& \left(1 - \frac{\mu \lambda_t \tau}{2} \right) \mathbb{E} \left[ \| \mathbf{w}_{t} - \mathbf{w}^* \|_2^2 \right]  + \lambda_t^2 \tau^2 G^2  \nonumber \\
			&+ \left(1 + \mu \lambda_t \right)\sum_{k = 1}^K \rho_k \sum_{l = 2}^{\tau} \mathbb{E} \left[ \|\mathbf{w}_{k, t}^{l} - \mathbf{w}_{t}\|_2^2 \right]  \nonumber \\
			&+ 2 L \lambda_t^2 \sum_{k = 1}^K \rho_k \sum_{l = 1}^{\tau} \mathbb{E} \left[ F_k(\mathbf{w}_{k, t}^{l}) - F_k^* \right]  \nonumber \\
			&- 2 \lambda_t \sum_{k = 1}^K \rho_k \sum_{l = 1}^{\tau} \mathbb{E} \left[ F_k(\mathbf{w}_{k, t}^{l}) - F_k(\mathbf{w}^*) \right],
	\end{align}
where $(a)$ is from \eqref{smoothness} and \eqref{lemma7_tri}, and $(b)$ is from the facts that $\mathbf{w}_{k, t}^{1} = \mathbf{w}_{t}$ and $\sum_{k = 1}^K \rho_k = 1$.
	
	Define $\psi_t = 2 \lambda_t (1 - L \lambda_t)$. Since $\lambda_t \leq \frac{1}{2 L \tau}$ and $\tau \geq 1$, we have $1 - L \lambda_t \geq 1 - \frac{1}{2 \tau} \geq \frac{1}{2}$ and thus $\lambda_t \leq \psi_t \leq 2\lambda_t$. Then, we have:
	\begin{align} \label{lemma7_decomp2_bound1}
			&2 L \lambda_t^2 \sum_{k = 1}^K \rho_k \sum_{l = 1}^{\tau} \mathbb{E} \left[ F_k(\mathbf{w}_{k, t}^{l}) - F_k^* \right]  \nonumber \\
			&- 2 \lambda_t \sum_{k = 1}^K \rho_k \sum_{l = 1}^{\tau} \mathbb{E} \left[  F_k(\mathbf{w}_{k, t}^{l}) - F_k(\mathbf{w}^*) \right]  \nonumber \\
			=& - \psi_t \sum_{k = 1}^K \rho_k \sum_{l = 1}^{\tau} \mathbb{E} \left[ F_k(\mathbf{w}_{k, t}^{l}) - F_k(\mathbf{w}^*) + F_k(\mathbf{w}^*) - F_k^* \right]  \nonumber \\
			&+ 2 \lambda_t \sum_{k = 1}^K \rho_k \sum_{l = 1}^{\tau} \mathbb{E} \left[ F_k(\mathbf{w}^*) - F_k^* \right]  \nonumber \\
			\overset{(c)}{=}& - \psi_t \sum_{k = 1}^K \rho_k \sum_{l = 1}^{\tau} \mathbb{E} \left[ F_k(\mathbf{w}_{k, t}^{l}) - F^* \right]  \nonumber \\
			&+ (2 \lambda_t - \psi_t) \sum_{l = 1}^{\tau} \left( F^* - \sum_{k = 1}^K \rho_k F_k^* \right)  \nonumber \\
			\overset{(d)}{=}& - \psi_t \sum_{k = 1}^K \rho_k \sum_{l = 1}^{\tau} \mathbb{E} \left[ F_k(\mathbf{w}_{k, t}^{l}) - F^* \right] + 2\lambda_t^2 \tau L  \Gamma,
	\end{align}
where $(c)$ is because $\sum_{k = 1}^K \rho_k F^* = F^* = \sum_{k = 1}^K \rho F_k(\mathbf{w}^*)$, and $(d)$ is because $\Gamma = \sum_{k = 1}^K \rho_k (F^* - F_k^*) = F^* - \sum_{k = 1}^K \rho_k F_k^*$.
	
	To bound the first term on the r.h.s. of \eqref{lemma7_decomp2_bound1}, we have
	\begin{align} \label{lemma7_decomp2_bound2}
			&\sum_{k = 1}^K \rho_k \sum_{l = 1}^{\tau} \mathbb{E} \left[ F_k(\mathbf{w}_{k, t}^{l}) - F^* \right]  \nonumber \\
			=& \sum_{k = 1}^K \rho_k \sum_{l = 1}^{\tau} \mathbb{E} \left[ F_k(\mathbf{w}_{k, t}^{l}) - F_k(\mathbf{w}_{t}) + F_k(\mathbf{w}_{t}) - F^* \right]  \nonumber \\
			\overset{(e)}{\geq}& \sum_{k = 1}^K \rho_k \sum_{l = 1}^{\tau} \mathbb{E} \left[ \langle \nabla F_k(\mathbf{w}_{t}), \mathbf{w}_{k, t}^{l} - \mathbf{w}_{t} \rangle + \frac{\mu}{2} \| \mathbf{w}_{k, t}^{l} - \mathbf{w}_{t} \|_2^2 \right]  \nonumber \\
			&+ \tau \mathbb{E} \left[ F(\mathbf{w}_{t}) - F^* \right]  \nonumber \\
			\overset{(f)}{\geq}& \sum_{k = 1}^K \rho_k \sum_{l = 1}^{\tau} \mathbb{E} \left[ - \frac{\lambda_t}{2} \| \nabla F_k(\mathbf{w}_{t})\|_2^2  - \frac{1}{2 \lambda_t} \| \mathbf{w}_{k, t}^{l} - \mathbf{w}_{t} \|_2^2 \right]  \nonumber \\
			&+ \frac{\mu}{2} \sum_{k = 1}^K \rho_k \sum_{l = 1}^{\tau} \mathbb{E} \left[ \| \mathbf{w}_{k, t}^{l} - \mathbf{w}_{t} \|_2^2 \right] + \tau \mathbb{E} \left[ F(\mathbf{w}_{t}) - F^* \right]  \nonumber \\
			\overset{(g)}{\geq}& - \sum_{k = 1}^K \rho_k \sum_{l = 1}^{\tau} \mathbb{E} \left[ \lambda_t L (F_k(\mathbf{w}_{t}) - F_k^*) \right] + \tau \mathbb{E} \left[ F(\mathbf{w}_{t}) - F^* \right] \nonumber \\
			&- \left(\frac{1}{2\lambda_t} - \frac{\mu}{2} \right) \sum_{k = 1}^K \rho_k \sum_{l = 1}^{\tau} \mathbb{E} \left[ \| \mathbf{w}_{k, t}^{l} - \mathbf{w}_{t} \|_2^2 \right],  \vspace{-0.3em}
	\end{align}
where $(e)$ is from the $\mu$-strong convexity of $F_k(\cdot)$, $(f)$ is from the AM-GM inequality, and $(g)$ is from \eqref{smoothness}. Substituting \eqref{lemma7_decomp2_bound2} into \eqref{lemma7_decomp2_bound1}, we have
	\begin{align} \label{lemma7_decomp2_bound3}
			&2 L \lambda_t^2 \sum_{k = 1}^K \rho_k \sum_{l = 1}^{\tau} \mathbb{E} \left[ F_k(\mathbf{w}_{k, t}^{l}) - F_k^* \right]  \nonumber \\
			&- 2 \lambda_t \sum_{k = 1}^K \rho_k \sum_{l = 1}^{\tau} \mathbb{E} \left[  F_k(\mathbf{w}_{k, t}^{l}) - F_k(\mathbf{w}^*) \right] \nonumber \\
			\leq& \psi_t \sum_{k = 1}^K \rho_k \sum_{l = 1}^{\tau} \mathbb{E} \left[ \lambda_t L (F_k(\mathbf{w}_{t}) - F^* + F^* - F_k^*) \right]  \nonumber \\
			&+ \psi_t \left(\frac{1}{2\lambda_t} - \frac{\mu}{2} \right) \sum_{k = 1}^K \rho_k \sum_{l= 1}^{\tau} \mathbb{E} \left[ \| \mathbf{w}_{k, t}^{l} - \mathbf{w}_{t} \|_2^2 \right]  \nonumber \\
			&- \psi_t \tau \mathbb{E} \left[ F(\mathbf{w}_{t}) - F^* \right] + 2 \lambda_t^2 \tau L \Gamma  \nonumber \\
			=& \psi_t (\lambda_t L - 1) \tau \sum_{k = 1}^K \rho_k \mathbb{E} \left[ F_k(\mathbf{w}_{t}) - F^* \right]  \nonumber \\
			&+ \left(2\lambda_t^2 \tau + \psi_t \lambda_t \tau \right) L \Gamma  \nonumber \\
			&+ \psi_t \left(\frac{1}{2\lambda_t} - \frac{\mu}{2} \right) \sum_{k = 1}^K \rho_k \sum_{l= 1}^{\tau} \mathbb{E} \left[ \| \mathbf{w}_{k, t}^{l} - \mathbf{w}_{t} \|_2^2 \right]  \nonumber \\
			\overset{(h)}{\leq}&- \frac{1}{2} \lambda_t \tau \mathbb{E} \left[ F(\mathbf{w}_{t}) - F^* \right] + 4 \lambda_t^2 \tau L \Gamma  \nonumber \\
			&+ \left(1 - \mu \lambda_t \right) \sum_{k = 1}^K \rho_k \sum_{l = 2}^{\tau} \mathbb{E} \left[ \| \mathbf{w}_{k, t}^{l} - \mathbf{w}_{t} \|_2^2 \right] ,
	\end{align}
where $(h)$ is due to the following facts, with each corresponding to one term:
	\begin{enumerate}
		\item $\lambda_t L - 1 \leq \frac{1}{2 \tau} - 1 \leq -\frac{1}{2}$, $\lambda_t \leq \psi_t \leq 2\lambda_t$, and $\sum_{k = 1}^K \rho_k (F_k(\mathbf{w}_{t}) - F^*) = F(\mathbf{w}_{t}) - F^* \geq 0$;
		
		\item $\Gamma \geq 0$ and $2\lambda_t^2 \tau + \psi_t \lambda_t \tau \leq 4 \lambda_t^2 \tau$;
		
		\item $L \geq \mu$, $\frac{1}{2\lambda_t} - \frac{\mu}{2} \geq L \tau - \frac{\mu}{2} \geq 0$, $\psi_t \big(\frac{1}{2\lambda_t} - \frac{\mu}{2} \big) \leq 2 \lambda_t \big(\frac{1}{2\lambda_t} - \frac{\mu}{2} \big) = (1 - \mu \lambda_t)$, and $\mathbf{w}_{k, t}^{1} = \mathbf{w}_{t}$.
	\end{enumerate}
Substituting \eqref{lemma7_decomp2_bound3} into \eqref{lemma7_decomp2}, we have
	\begin{align}  \label{lemma7_eq}
			&\mathbb{E} \left[ \| \mathbf{v}_{t + 1} - \mathbf{w}^* \|_2^2 \right]  \nonumber \\
			\leq& \left(1 - \frac{\mu \lambda_t \tau}{2} \right) \mathbb{E} \left[ \| \mathbf{w}_{t} - \mathbf{w}^* \|_2^2 \right] - \frac{1}{2} \lambda_t \tau \mathbb{E} \left[ F(\mathbf{w}_{t}) - F^* \right]  \nonumber \\
			&+ \lambda_t^2 \tau^2 G^2 + 2 \sum_{k = 1}^K \rho_k \sum_{l = 2}^{\tau} \mathbb{E} \left[ \|\mathbf{w}_{k, t}^{l} - \mathbf{w}_{t}\|_2^2 \right] + 4 \lambda_t^2 \tau L \Gamma.
	\end{align}
Plugging \eqref{lemma7_eq} and $\mathbb{E} \left[ \langle \hat{\mathbf{r}}_t - \mathbf{r}_t, \mathbf{v}_{t + 1} - \mathbf{w}^* \rangle \right] = 0$ into \eqref{lemma3_proof_eq2} gives the result in \eqref{lemma3_eq}.

\bibliography{references}
\end{document}